\documentclass{aa}  

\makeatletter

\let\AA@old@journalname\aa@journalname
\let\AA@old@manuscriptname\aa@manuscriptname
\let\AA@old@AALogo\AALogo
\let\AA@old@today\today
\newlength\AA@old@fboxrule
\newlength\AA@old@fboxsep

\newcommand*\AA@disableTitleBanner{%
	\renewcommand*\aa@journalname{}
	\renewcommand*\aa@manuscriptname{}
	\renewcommand*\AALogo{}
	\def\today{}
	\setlength{\AA@old@fboxrule}{\fboxrule}
	\setlength{\AA@old@fboxsep}{\fboxsep}%
	\setlength{\fboxrule}{0pt}%
	\setlength{\fboxsep}{0pt}%
}

\newcommand*\AA@restoreTitleBanner{%
	\let\aa@journalname\AA@old@journalname
	\let\aa@manuscriptname\AA@old@manuscriptname
	\let\AALogo\AA@old@AALogo
	\let\today\AA@old@today
	\setlength{\fboxrule}{\AA@old@fboxrule}%
	\setlength{\fboxsep}{\AA@old@fboxsep}%
}

\let\AA@old@maketitle\maketitle
\renewcommand*\maketitle{%
	\AA@disableTitleBanner
	\AA@old@maketitle
	\AA@restoreTitleBanner
}

\makeatother

\usepackage{fancyhdr}
\fancypagestyle{plain}{%
	\fancyhf{}                 
	\fancyfoot[C]{\thepage}    
	
}
\usepackage[utf8]{inputenc}
\usepackage[varg]{txfonts}  
\usepackage{bm}
\usepackage{amssymb}
\usepackage{amsmath}
\usepackage{nicefrac}
\usepackage{mathrsfs}
\usepackage{graphicx}
\usepackage{tabularx}
\usepackage{enumitem}
\usepackage{color}
\usepackage{subcaption}
\usepackage{hyperref}
\usepackage{booktabs}
\usepackage{multirow}
\usepackage{array}
\usepackage{upgreek}
\usepackage{appendix}
\makeatletter
\AtBeginDocument{%
	\geometry{
		hmargin=2cm
	}%
}

\hypersetup{
	colorlinks = true,
	linkcolor = blue,
	citecolor = blue,
	filecolor = blue,
	urlcolor = blue
}

\def\apjl{ApJL}
\def\apjs{Astrophys. J. Supp.}

\def\aap{Astron. Astrophys.}

\newcolumntype{K}[1]{>{\centering\arraybackslash}p{#1}}

\let\OLDthebibliography\thebibliography
\renewcommand\thebibliography[1]{
	\OLDthebibliography{#1}
	\setlength{\parskip}{0pt}
	\setlength{\itemsep}{0pt plus 0.3ex}
}

\allowdisplaybreaks

\titlerunning{Accretion-disk formation in gaseous star clusters}

\title{Accretion-disk formation around orbiting stellar black holes in gaseous star clusters}

\author{Zacharias Roupas\inst{1,2}\thanks{\email{zacharias.roupas@unimib.it}}}

\institute{Dipartimento di Fisica ``G. Occhialini'', 
	Universit\'a degli Studi di Milano-Bicocca, Piazza della Scienza 3, 20126 Milano, Italy
	\and
	Istituto Nazionale di Fisica Nucleare (INFN), Sezione di Milano-Bicocca, 
	Piazza della Scienza 3, 20126 Milano, Italy
}

\date{}

\abstract{
	We consider low-mass black holes (BHs) moving in regular orbits in the cores of non-rotating gaseous star clusters, representative of proto-stellar clusters or the centers of protogalaxies. We argue that as the BH's sphere of influence -- the Bondi sphere -- is advected along the BH trajectory, the transverse velocity shear between the inner and outer hemispheres injects angular momentum, driving the formation of an accretion disk. Coriolis forces oppose angular momentum injection, delaying disk formation but not preventing it. The disk lies in the BH orbital plane and is counter-rotating with respect to the orbital BH motion. We verify this picture with 2D and 3D hydrodynamic simulations in the non-inertial frame of the orbiting BH. We find that the disk-formation timescale following a disruption event is of order the Bondi crossing timescale, $\tau_{\rm d} \sim R_{\rm B}/V_{\bullet}$, and that the disk radius is of order $R_{\rm d} \sim \omega_{\bullet}^2 R_{\rm B}^4/ G m_{\bullet} $, set by the circularization radius of gas captured in the Bondi sphere. For the case of a BH with $m_{\bullet} = 50\,{\rm M}_\odot$, inside the core of a typical compact proto-stellar cluster, these values read $\tau_{\rm d} \sim 0.1 P_{\rm orbit}$ and $R_{\rm d} \sim 10^{-3} R_{\rm B}$.
}

\begin{document}
	
	\maketitle
	
	\section{Introduction}\label{sec:intro}
	
	The origin and form of the mass and spin distributions of stellar black holes (BHs) are central open questions in gravitational-wave (GW) astronomy \citep{2021ApJ...913L...7A,2023PhRvX..13a1048A}, especially in view of the growing body of GW data reported by \citet{2025arXiv250818083T}.
	We focus here on stellar BHs residing in gaseous star clusters \citep{2013MNRAS.429.2997L,2019A&A...632L...8R, 2021A&A...646A..20R, 2025A&A...702A.208R, 2026Parti...9...18R, 2026A&A...709A...5R}. This setting is relevant to the so-called dynamical channel of binary black hole (BBH) mergers
		 \citep{2002MNRAS.330..232C, 2006ApJ...637..937O, 2016PhRvD..93h4029R, 2016ApJ...831..187A, 2017PhRvD..95l4046G, 2019A&A...621L...1R,2021MNRAS.505..339M,2025PhRvL.134a1401A,2026arXiv260407456G}. 
		 We do not consider stellar BHs embedded in active galactic nucleus (AGN) disks \citep[e.g.][]{2012MNRAS.425..460M, 2017MNRAS.464..946S, 2017ApJ...835..165B, 2020ApJ...901L..34Y, 2020ApJ...898...25T, 2026ApJ...996L..44B}, where the ambient gas possesses coherent angular momentum. We instead restrict ourselves to non-rotating gas configurations.
		 
		An accretion disk can modify the BH spin through the transfer of angular momentum carried by the gas. Predicting BH spin values is important for distinguishing between different BBH-merger channels \citep{2016ApJ...832L...2R,2021ApJ...915...56G,2022PhR...955....1M,2025ApJ...994L..37K,2026A&A...708A..62W,2026PhRvD.113d3048B,2026A&A...709A...5R}. This motivates us to understand whether and how an accretion disk may form around a stellar BH, and to identify its physical properties.
	
	In the case of a supermassive black hole (SMBH) residing at the gravitational center of a gaseous system, the formation of an accretion disk is naturally driven by angular momentum carried by the inflowing gas within the SMBH's gravitational sphere of influence \citep{1969Natur.223..690L,1981ARA&A..19..137P}. 
	Irrespective of the mechanism that provides the angular momentum \citep[e.g.][]{1989Natur.338...45S,2004MNRAS.354..292K,2006MNRAS.370..289B,2006MNRAS.371.1813L,2010MNRAS.407.1529H,2010Natur.466.1082M,2011MNRAS.415.1027H,2013MNRAS.432.3401G}, the SMBH occupies a privileged position at the center of the gravitational potential. Thus, inflowing material with sufficient residual angular momentum forms a disk provided that the circularization radius exceeds the innermost stable circular orbit (ISCO).
	
		The situation is very different for a stellar-mass BH moving rapidly inside a gaseous star cluster. The gas environment in the immediate vicinity of the BH changes as the BH moves within the ambient gas. It is not a priori clear whether a persistent accretion disk would form, especially in non-rotating or weakly rotating star clusters. Although we are primarily concerned with gaseous proto-stellar clusters, plausible precursors of present-day globular clusters (GCs) \citep{2024Natur.632..513A}, we note that Gaia-era studies suggest that the majority of Milky Way GCs are pressure-supported systems, and either non-rotating or weakly to modestly rotating \citep{2019MNRAS.485.1460S,2021MNRAS.505.5978V}. 
		The formation of disks around stellar BHs due to general gas velocity gradients has been studied with hydrodynamic simulations \citep{1995A&A...295..108R,1997A&A...317..793R,2019MNRAS.488.5162X,2025ApJ...979...61T}. These studies, however, do not identify the astrophysical origin of such gradients in star clusters. 
		
		Recently, we proposed a disk-formation mechanism \citep{2026A&A...709A...5R} based on the physical argument that the azimuthal velocity of the BH itself generates gas velocity gradients.
		Here, we study this disk-formation mechanism in detail. The velocity shear between the inner and outer hemispheres of the Bondi sphere supplies angular momentum as the sphere is advected along the BH trajectory. We derive analytically the kinematic equations of the gas in the non-inertial frame of the BH. We then verify accretion-disk formation with 3D and 2D hydrodynamic simulations and estimate the disk size and formation timescale. 
		In Sect.~\ref{sec:kin}, we derive the kinematic equations. In Sect.~\ref{sec:sims} we present the simulation results, and we conclude in Sect.~\ref{sec:conclusion}.

	\section{Kinematics}\label{sec:kin}
	
\begin{figure}[tbp]
		\centering
		\includegraphics[width=0.6\columnwidth]{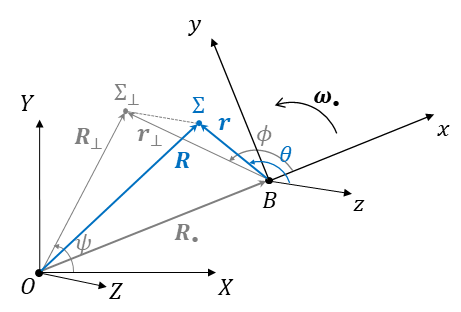}
		\caption{The rotating coordinate system B$xyz$ attached to the BH orbit, and the inertial coordinate system O$XYZ$ with origin the center of the cluster. The $z$-axis and $Z$-axis are aligned.}
		\label{fig:coord}
	\end{figure}

	We define in Figure \ref{fig:coord} the non-inertial, rotating frame $Bxyz$ of a BH at point $B$ orbiting with angular velocity $\bm{\omega}_{\bullet}(t)$ about $OXYZ$, the inertial frame in the center of the gaseous cluster, assuming $z$-axis and $Z$-axis are aligned. This assumption of parallelism between $xy$ and $XY$ planes entails no loss of generality provided we also assume negligible gas rotation about the cluster center.
	We identify $\hat{x} \equiv \hat{R}_{\bullet}$ and $\hat{y} \equiv \hat{\psi}$.
	The polar coordinates in the $XY$ plane are $(R_{\perp}, \psi)$ and in the $xy$ plane they are $(r_\perp,\phi)$. Correspondingly, the spherical coordinates in the $Bxyz$ frame are $(r,\phi,\theta)$ with $r_{\perp} = r\sin\theta$, $\theta \in {[0,\pi)}$ and $\phi \in {[0,2\pi)}$.
	
	\subsection{Apparent velocity}\label{sec:vel}

	The apparent velocity $\bm{v}_{\Sigma}$ of a point $\Sigma$ in the BH frame, $Bxyz$, is
		\begin{equation}
			\label{eq:v_Sigma}
			\bm{v}_{\Sigma} = \bm{v}^{(O)}_{\Sigma} -  \bm{V}_{\bullet} - \bm{\omega}_{\bullet} \times \bm{r} ,
		\end{equation}
		where in general all quantities are time-dependent.
		The last term implies rotation counter to the BH orbital rotation.
		The term $\bm{v}^{(O)}_{\Sigma}$ is the motion of the gas in the cluster frame $OXYZ$, for example it is $\bm{v}^{(O)}_{\Sigma} = 0$ for a hydrostatic equilibrium and $\bm{v}^{(O)}_{\Sigma} = \bm{\omega}_{\rm gas} \times \bm{R}$ for rotating gas.
		 We assume $\bm{v}^{(O)}_{\Sigma} = 0$.
		 The BH velocity in the $OXYZ$ frame is (we do not use a superscript for ${\bm V}_{\bullet}$ to ease notation, since it is non-zero only in the $OXYZ$ frame)
		 \begin{equation}
		 	\bm{V}_{\bullet} = V_{\bullet,R}\hat{R}_{\bullet} + V_{\bullet,\psi}\hat{\psi}_{\bullet} =
		 	V_{\bullet,R}\hat{x} + V_{\bullet,\psi}\hat{y}.
		 \end{equation}
		 We also have, $\bm{\omega}_{\bullet}\times \bm{r} = \omega_{\bullet} r \sin\theta \,\hat{\phi}$.
		 We get in spherical coordinates in the $Bxyz$ frame:
		 \begin{align}
		 	\bm{v}_{\Sigma} = &-\left(V_{\bullet,R}\cos\phi + V_{\bullet,\psi}\sin\phi\right)\left( \sin\theta\, \hat{r} + \cos\theta \,\hat{\theta} \right)
		 	\nonumber \\
		 	\label{eq:v_Sigma_an}
		 	&- \left(\omega_{\bullet} r \sin\theta - V_{\bullet,R}\sin\phi + V_{\bullet,\psi}\cos\phi\right) \hat{\phi}
		 	.
		 \end{align}
		 This is the velocity field in the BH moving frame of reference. 
		 
		 \subsection{BH equations of motion}\label{sec:eom}
		 
		 The BHs get rapidly segregated in the core of a dense gaseous star cluster \citep{Spitzer_1987degc, 2014MNRAS.441..919L}, where the gas and stellar density profiles are nearly flat. Therefore, we assume $\rho_{\rm cl} \approx {\rm const}$, where $\rho_{\rm cl}$ denotes the total mass density of the cluster including both gas and star components.
		 We discuss further the uniform density assumption in section \ref{sec:theory} below.
		  Besides the physical motivation, isolating the velocity gradients allows us to quantify their sole effects on disk formation. The BH is subject to the gravitational acceleration
		 \begin{equation}
		 	\bm{a}_{\rm G} = - \Omega^2 R_{\bullet} \,\hat{R}_{\bullet}, \; \Omega \equiv \sqrt{\frac{4\pi G\rho_{\rm cl}}{3}}.
		 \end{equation}
		 The polar equations in the $XY$ plane are
		 \begin{align}
		 	\ddot{R}_{\bullet} - R_{\bullet} \dot{\psi}^2 &= -\Omega^2 R_{\bullet}, 
		 	\\
		 	R_{\bullet} \ddot{\psi} + 2\dot{R}_{\bullet} \dot{\psi} &= 0 .
		 \end{align}
		 The latter is just the angular momentum conservation $\ell_{\bullet} = R_{\bullet}^2 \dot{\psi} = {\rm const}$. We differentiate twice the Cartesian coordinates 
		 \begin{align}
		 	X_{\bullet}(t) &= R_{\bullet}(t) \cos\psi (t), 
		 	\\
		 	Y_{\bullet}(t) &= R_{\bullet}(t) \sin\psi (t), 
		 \end{align}		 
		 and substitute the polar equations of motion to get the decoupled oscillators for the Cartesian coordinates
		 \begin{align}
		 	\ddot{X}_{\bullet}(t) + \Omega^2 X_{\bullet}(t) &= 0, 
		 	\\
		 	\ddot{Y}_{\bullet}(t) + \Omega^2 Y_{\bullet}(t) &= 0.
		 \end{align}
		 The solution for a bound system is an ellipse with the center of the gravitational field of the cluster $O$ identified with the center of the ellipse. Assuming the $OXYZ$ frame is oriented such that the major ellipse's axis lies along the $X$-axis the solution is
		 \begin{align}
		X_{\bullet}(t) &= A \cos\left(\Omega\, t + \xi_0 \right), 
		\\
		Y_{\bullet}(t) &= B \sin\left(\Omega\, t + \xi_0\right),
	\end{align}
	where we denote the semi-major axis $A$ and the semi-minor axis $B$.
	This ellipse is actually a closed Lissajous orbit.
	The initial phase $\xi_0$ is calculated from the initial condition $\psi (t=0) = \psi_0$ 
	\begin{equation}
		\xi_0 = {\rm atan2} \left(\frac{Y_0}{B},\frac{X_0}{A} \right) = {\rm atan2} \left(A\sin\psi_0,B\cos\psi_0 \right).	
	\end{equation}
	
	Assuming that the semi-minor axis $B$ and the eccentricity $e$ are known, the semi-major axis is
	\begin{equation}
		A = \frac{B}{\sqrt{1-e^2}}.
	\end{equation}
	We have $\ell_{\bullet} = X\dot{Y} - Y\dot{X}$ which gives
	\begin{equation}
		\ell_{\bullet} = \Omega B A.
	\end{equation}
	The angular velocity is 
	\begin{equation} 
				 	\label{eq:omega_BH}
		\omega_{\bullet}(t) = \frac{\ell_{\bullet}}{R_{\bullet}(t)^2} 
	\end{equation} 
	and the velocity components $V_{\bullet,R} = \dot{R}_{\bullet}$, $V_{\bullet,\psi} = \omega_{\bullet} R_{\bullet}$ 
		 \begin{align}
		 	\label{eq:V_BH_R}
		 	V_{\bullet,R}(t) &= - 
		 	\frac{1}{2} \Omega (B^2 - A^2) \frac{\sin\left( 2(\Omega\, t + \xi_0) \right)}{R_{\bullet}(t)},
		 	\\
		 	\label{eq:V_BH_xi}
		 	V_{\bullet,\psi}(t) &= \frac{\ell_{\bullet}}{R_{\bullet}(t)},
			\\
		 	\label{eq:R_BH}
			R_{\bullet}(t) &= \sqrt{X_{\bullet}(t)^2 + Y_{\bullet}(t)^2},
			\\
		 	\label{eq:xi_BH}
		 	\psi (t) &= {\rm atan2} (Y_{\bullet}(t),X_{\bullet}(t)) .
		 \end{align} 
		
		\begin{figure}[tbp]
			\centering
			\begin{subfigure}{\columnwidth}
				\centering
				\includegraphics[width=0.9\columnwidth]{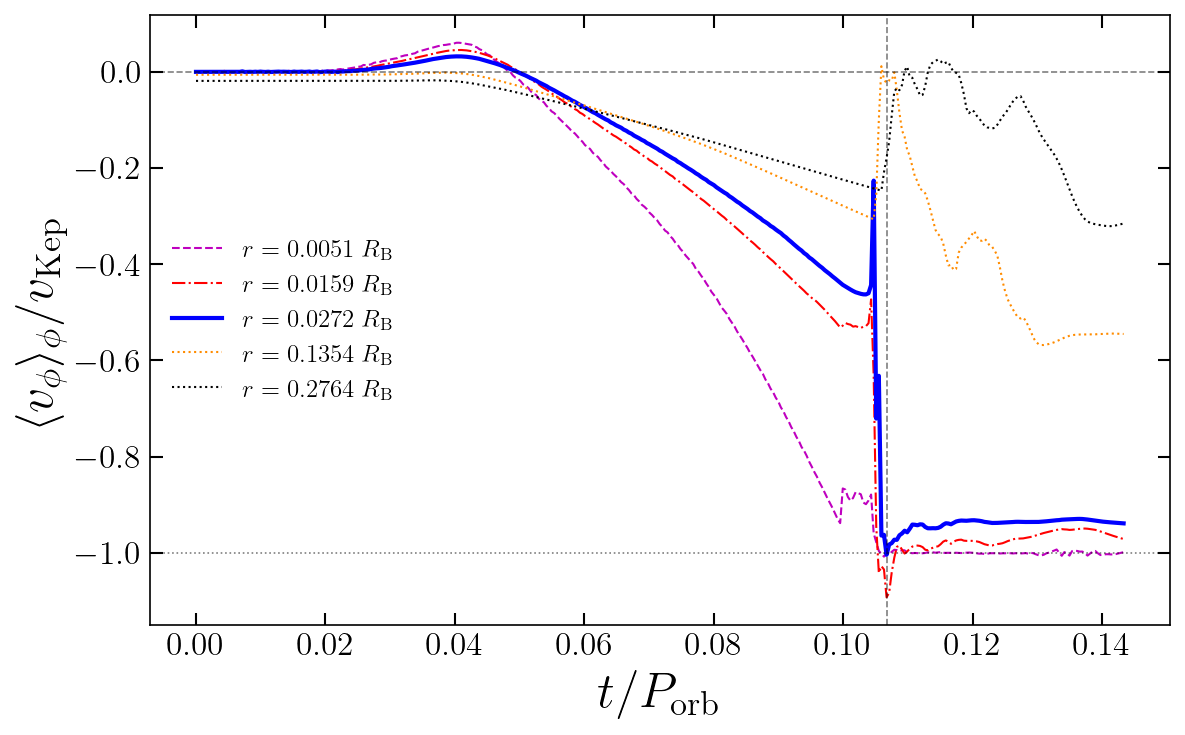}
				\caption{}
				\label{fig:vphi_time_2D}
			\end{subfigure}
			\begin{subfigure}{\columnwidth}
				\centering
				\includegraphics[width=0.9\columnwidth]{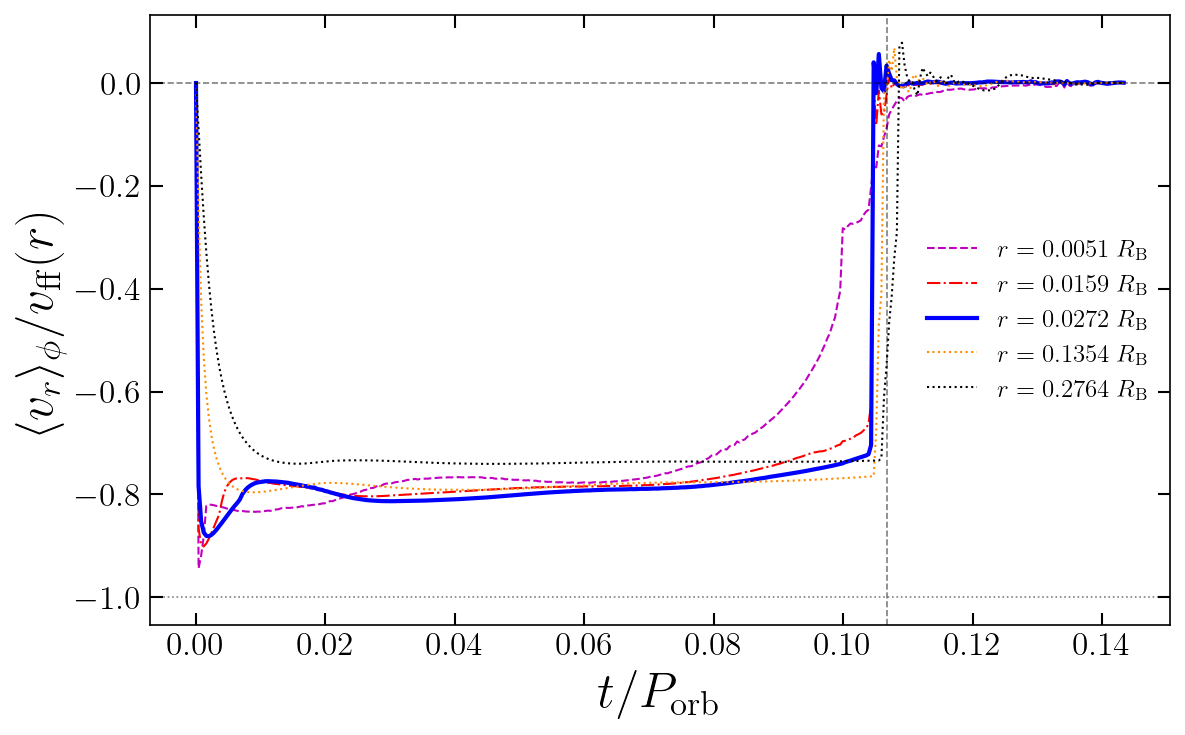}
				\caption{}
				\label{fig:vr_time_2D}
			\end{subfigure}
			\caption{A 2D simulation run for $m_{\bullet} = 50{\rm M}_{\odot}$, $\rho_{\rm cl} = 10^7{\rm M}_{\odot}/{\rm pc}^3$ and circular BH orbit. The azimuthal velocity averaged over $\phi$ normalized by the Keplerian velocity (panel (a)) and the radial velocity averaged over $\phi$ normalized by the free-fall velocity (panel (b)) with respect to time for several radii. The vertical gray dotted line marks the disk formation timescale $\tau_{\rm d}$ and the solid blue line represents the evolution at the disk radius $R_{\rm d}$.}
			\label{fig:v_time_2D}
		\end{figure}
			
		\subsection{Non-inertial acceleration}\label{sec:acc}
		
		The acceleration in the BH frame, $Bxyz$, is
		\begin{equation}
			\label{eq:a_Sigma}
			\bm{a}_{\Sigma} = \bm{a}_{\Sigma}^{(O)} - \bm{a}_{\bullet} - \bm{\omega}_{\bullet} \times(\bm{\omega}_{\bullet} \times \bm{r})
			- 2 \bm{\omega}_{\bullet} \times \bm{v}_{\Sigma}
			- \dot{\bm{\omega}}_{\bullet}\times \bm{r},
		\end{equation}
		where 
		\begin{equation}
			\bm{a}_{\Sigma}^{(0)} = - \Omega^2 (\bm{R}_{\bullet} + \bm{r}) - G\frac{m_{\bullet}}{r^2}\hat{r}
		\end{equation}
		and $\bm{a}_{\bullet} = - \Omega^2 \bm{R}_{\bullet}$.
		The tidal, centrifugal and Coriolis acceleration are, respectively,
		\begin{align}
			\bm{a}_{\rm T} &= - \Omega^2 r\,\hat{r}, 
			\\
			\bm{a}_{\rm cent} &\equiv - \bm{\omega}_{\bullet} \times(\bm{\omega}_{\bullet} \times \bm{r})
			= \omega_{\bullet}^2 r \sin\theta\left(\sin\theta \,\hat{r} + \cos\theta\,\hat{\theta}\right),
			\\
			\label{eq:a_cor}
			\bm{a}_{\rm cor} &\equiv - 2 \bm{\omega}_{\bullet} \times \bm{v}_{\Sigma} = 2\omega_{\bullet} v_{\phi}\sin\theta\,\hat{r} \,- 
			\nonumber \\
			&\quad - 2\omega_{\bullet} \left(v_r\sin\theta + v_{\theta}\cos\theta\right)\,\hat{\phi} 
			+ 2\omega_{\bullet} v_{\phi}\cos\theta\,\hat{\theta}
			.
		\end{align}
		Therefore, the acceleration of gas at point $\Sigma$,  Eq.~(\ref{eq:a_Sigma}), 
		is in spherical coordinates in the BH frame, $Bxyz$, 
		\begin{equation}
			\bm{a}_{\Sigma} = a_r\, \hat{r} +  a_{\theta}\,\hat{\theta} + a_{\phi}\,\hat{\phi}
		\end{equation}
		with
		\begin{align}
			\label{eq:a_r}
			a_r &= -G\frac{m_{\bullet}}{r^2} - \Omega^2 r + \omega_{\bullet}^2 r\sin^2\theta + 2 \omega_{\bullet}v_{\phi}\sin\theta
			\\
			\label{eq:a_th}
			a_{\theta} &= \omega_{\bullet}^2 r \sin\theta\cos\theta  + 2\omega_{\bullet} v_{\phi}\cos\theta,
			\\
			\label{eq:a_phi}
			a_{\phi} &= - \dot{\omega}_{\bullet} r\sin\theta - 2 \omega_{\bullet} \left( v_r \sin\theta + v_{\theta}\cos\theta \right),
		\end{align}
		where
		\begin{equation}
			\dot{\omega}_{\bullet} = - 2\ell_{\bullet} \frac{V_{\bullet,R}}{R_{\bullet}^3} .
		\end{equation}
			 					 
		 \subsection{Theoretical analysis}\label{sec:theory}
		 
		 Equation (\ref{eq:v_Sigma_an}) represents the gas velocity with respect to the BH outside the BH's gravitational sphere of influence. Let us assume it is the boundary velocity at the Bondi sphere 
		 \begin{equation} 
		 	R_{\rm B} = \frac{2 G m_{\bullet}}{c_s^2 + V_{\bullet}^2}.
		\end{equation}
		 For a circular BH orbit, $\omega_{\bullet} = \Omega = {\rm const}$ and $V_{\bullet} = \Omega R_{\bullet} = {\rm const}$, we get that gas with average azimuthal velocity,
		 $\langle v_{\phi, {\rm in}} \rangle_{\phi} \approx \int_0^{2\pi} v_{\phi}(R_{\rm B},\theta,\phi)d\phi / 2\pi = - \Omega R_{\rm B} \sin\theta$, is injected into the Bondi sphere. This comes only from the $\omega_{\bullet} r$ term of Eq.~(\ref{eq:v_Sigma_an}), since both $V_{\bullet}$ terms are zero over $\phi$ integration. Since $\sin\theta$ is positive, the azimuthal velocity drives a rotation of the gas counter to the orbital rotation of the BH.
		 The $v_{\phi, {\rm in}}$ represents the velocity shear between the inner and outer hemispheres of the Bondi sphere. 
		 This may be understood as follows. The relative velocity between gas and BH of the inner-most boundary point of the Bondi sphere (inner hemisphere) on the orbital plane is ${\bm v}_1 = \Omega (R_{\bullet} - R_{\rm B})\,\hat{\phi}$ and at the outer-most point (outer hemisphere) is ${\bm v}_2 = - \Omega (R_{\bullet} + R_{\rm B}) \,\hat{\phi}$. 
		 There is therefore an average velocity shear between the two points, $({\bm v}_1 + {\bm v}_2)/2 = - \Omega R_{\rm B}\,\hat{\phi}$.
		 
		 Let us now calculate the injection of the $z$-component angular momentum over the whole Bondi shell. For a circular BH orbit, taking the average of $v_{\phi}$ over $\phi$, and averaging also over the $z$-planes we get 
		 \begin{equation}
		 	j_{z,{\rm in}}^{(\rm shear)}
		 	 \approx  \frac{1}{4\pi}\iint R_{\rm B} v_{\phi}(R_{\rm B},\theta,\phi) \sin^2\theta \,d\theta \,d\phi
		 	\label{eq:jin_circ}
		 	= - \frac{2}{3} \Omega R_{\rm B}^2 ,
		 \end{equation}
		 where the minus sign represents counter-rotation with respect to the BH orbital motion. 
		 For an elliptic orbit we would expect a different geometric factor of order one, so that in general we expect as an order of magnitude $j_{z,{\rm in}} \sim - \Omega R_{\rm B}^2$.
		 		 Identifying the disk radius with the circularization radius for this angular momentum injection,
		 $R_{\rm d} \equiv j_{\rm in}^2 / (G m_{\bullet})$, 
		 we have that
		 \begin{equation}\label{eq:R_d}
		 	R_{\rm d} \sim \lambda_{\rm R} 
		 	R_{\rm circ} 
		 	\, , \quad
		 	R_{\rm circ} \equiv \frac{\Omega^2 R_{\rm B}^4}{G m_{\bullet}},
		 \end{equation}
		 where the prefactor $\lambda_{\rm R}$ is of order one and shall be estimated empirically.
		 
		 The explicit formula  Eq.~(\ref{eq:jin_circ}) allows us to further justify the uniform density assumption. 
		 The maximum density difference, $\Delta \rho_{\rm max}$, is between the innermost ($R_{\bullet} - R_{\rm B}$) and the outermost ($R_{\bullet} + R_{\rm B}$) boundary of the Bondi sphere. To first order in $R_{\rm B} / R_{\bullet}$, it is $\Delta \rho_{\rm max} \approx 2\rho^{\prime}(R_{\bullet}) R_{\rm B}$. Expanding also $|\bm{R}_{\bullet}  + R_{\rm B}\hat {r}|$, we get that at a point on the boundary of the Bondi sphere the density is $\rho(\theta, \phi) \approx \rho(R_{\bullet}) + \frac{\Delta \rho_{\rm max}}{2}\sin\theta\cos\phi$, that is $\Delta \rho(\theta, \phi) \approx \frac{\Delta \rho_{\rm max}}{2}\sin\theta\cos\phi$. We can now calculate the injected specific angular momentum induced by the density gradient by direct integration
		 \begin{align}
		 	j_{z,{\rm in}}^{(\rho)} 
		 	&\approx
		 	\frac{1}{4\pi} \iint 
		 	\frac{\Delta\rho (\theta,\phi)}{\rho(R_{\bullet})} R_{\rm B} v_{\phi}(R_{\rm B},\theta,\phi) \sin^2\theta \,d\theta \,d\phi
		 	\nonumber \\
		 	&=   -\frac{1}{3}\frac{\rho^{\prime}(R_{\bullet})}{\rho(R_{\bullet})} R_{\bullet} \Omega R_{\rm B}^2 ,
		 \end{align}
		 where we substituted the boundary $v_{\phi}(R_{\rm B},\theta,\phi)$ from Eq.~(\ref{eq:v_Sigma_an}) for a circular orbit, and performed the angular integrations explicitly.
		 As a conservative estimate, let us adopt a Plummer density profile, whose core gradient is steeper than that of a King profile.
		 The BHs are driven inside the core of the cluster by mass segregation \citep{Spitzer_1987degc}. 
		 We estimate below that $R_{\bullet} / r_{c} = \mathcal{O}(10^{-2})$ for our system (see also \citet{2021A&A...646A..20R}). 
	For the Plummer density, $\rho(r) \propto (1 + (r/r_c)^2)^{-5/2} $, we have to first order in $R_{\bullet}/r_c$ that $\rho^{\prime}(R_{\bullet})/\rho(R_{\bullet}) \approx -5 R_{\bullet}/r_c^2$, which gives
	\begin{equation}
		j_{z,{\rm in}}^{(\rho)} \approx 
		\frac{5}{3} \left(\frac{R_{\bullet}}{r_c} \right)^2 \Omega R_{\rm B}^2 ,
		\label{eq:jin_rho}
	\end{equation}
	The ratio between shear and density-gradient specific angular momentum follows directly
	\begin{equation}
		\left|\frac{j_{z,\rm in}^{(\rho)}}{j_{z,\rm in}^{({\rm shear})}}\right|
		\approx \frac{5}{2}\left(\frac{R_{\bullet}}{r_c}\right)^2 \sim 10^{-4}.
	\end{equation}
Therefore, deep inside the core of the cluster where our BHs reside, the specific angular momentum induced by density gradients is negligible compared to the shear-induced one. 
		 
		  \begin{figure*}[tbp]
		 	\centering
		 	\includegraphics[width=0.65\columnwidth]{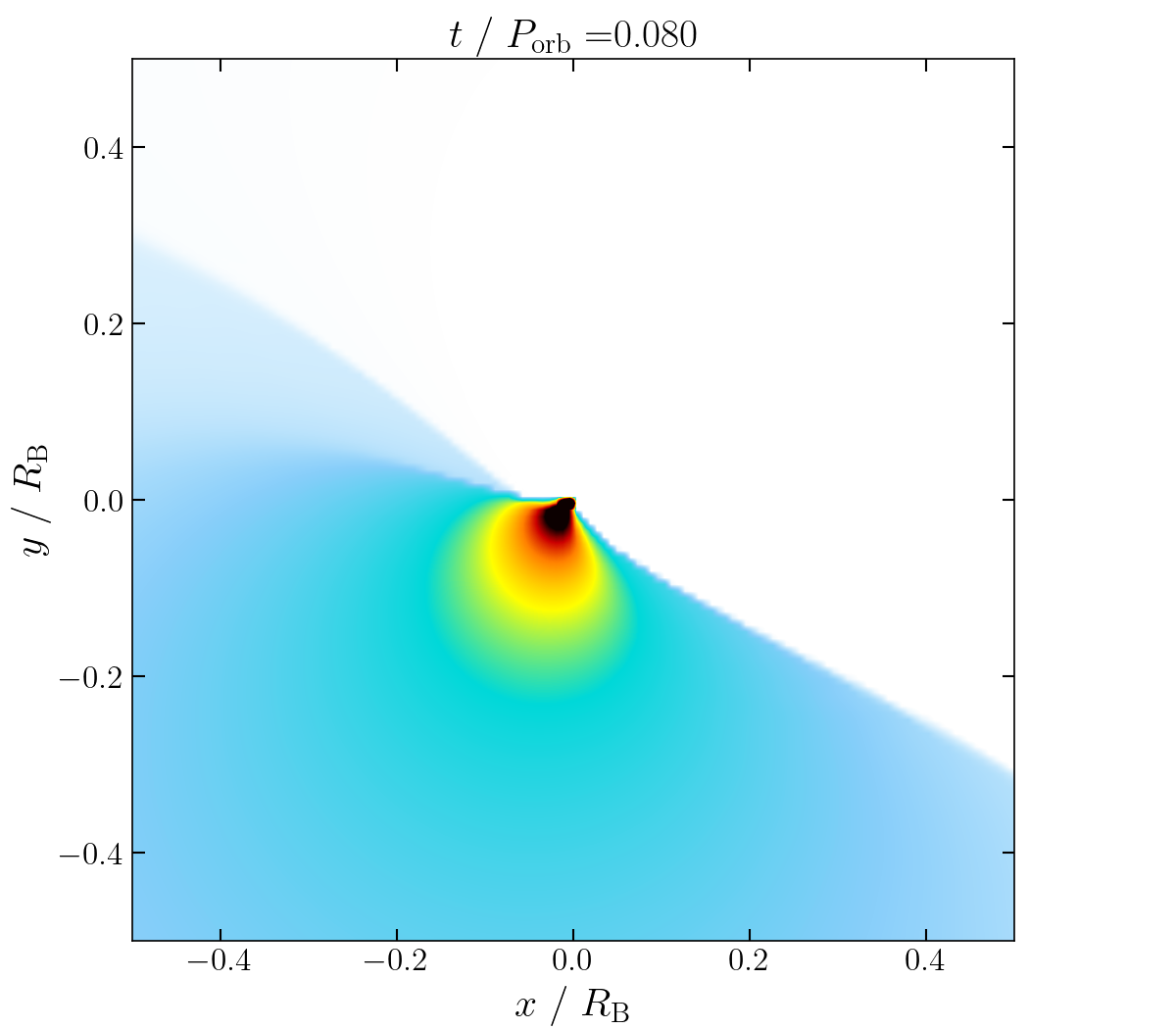}
		 	\includegraphics[width=0.65\columnwidth]{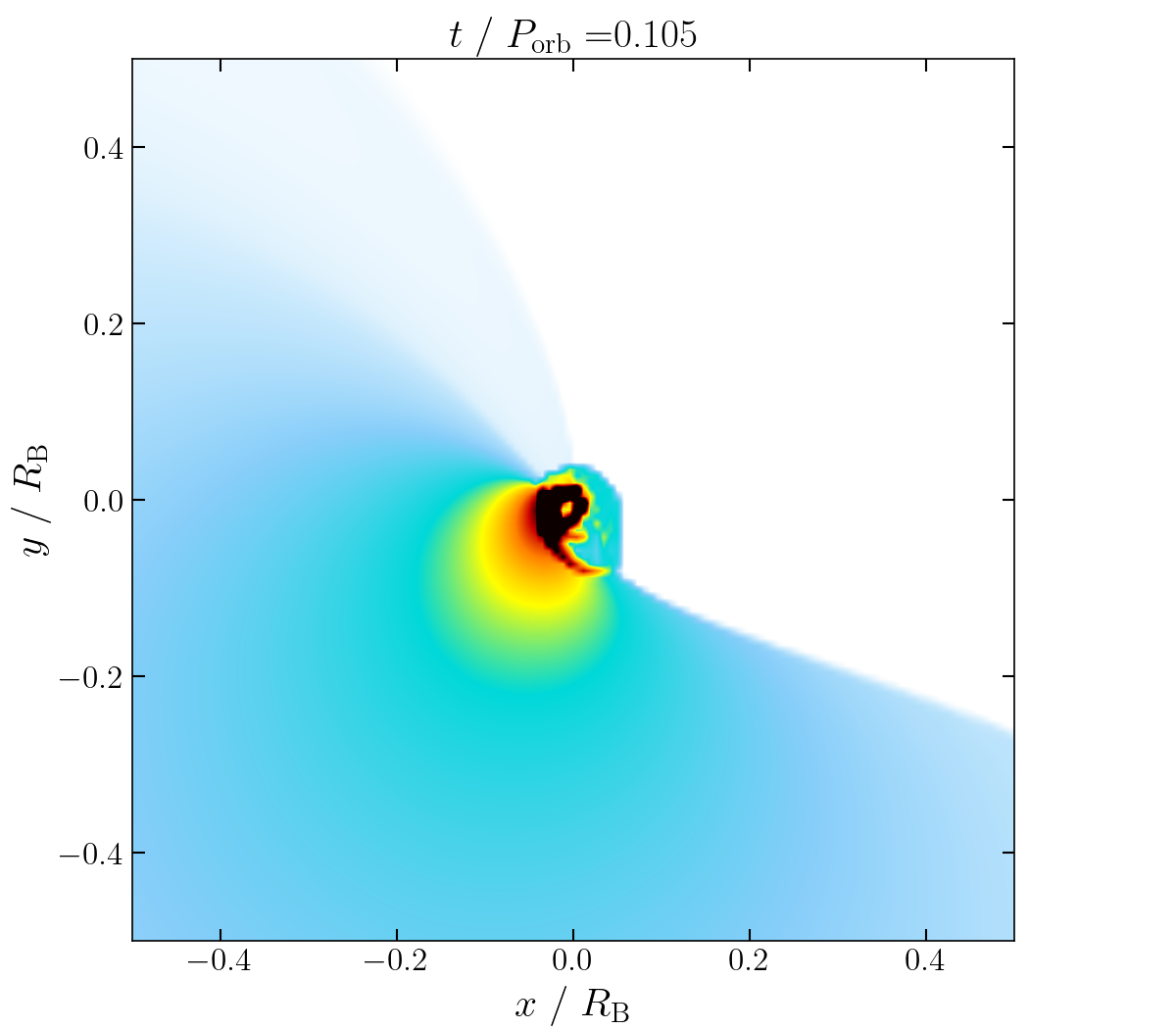}
		 	\includegraphics[width=0.65\columnwidth]{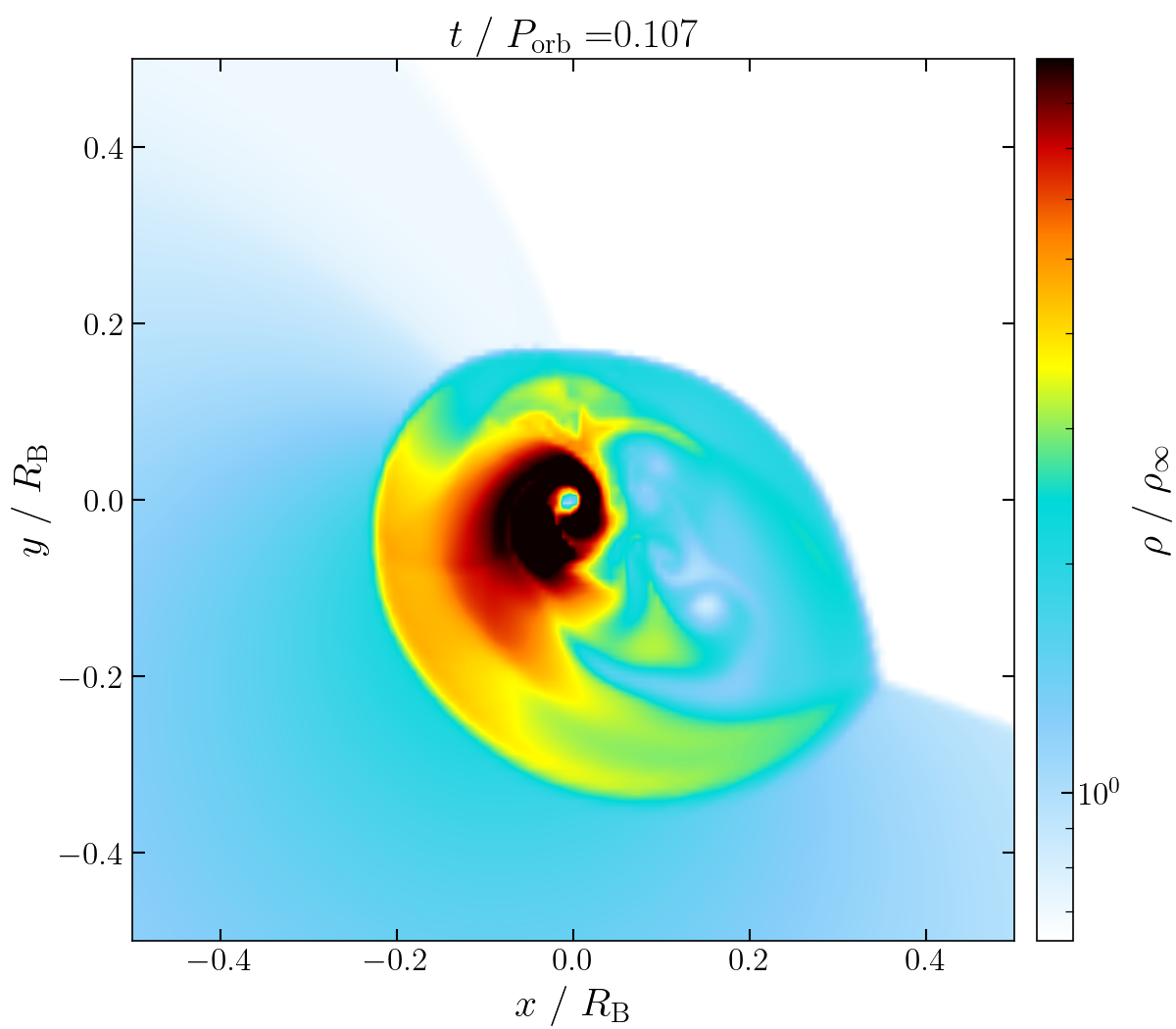}
		 	\\
		 	\includegraphics[width=0.65\columnwidth]{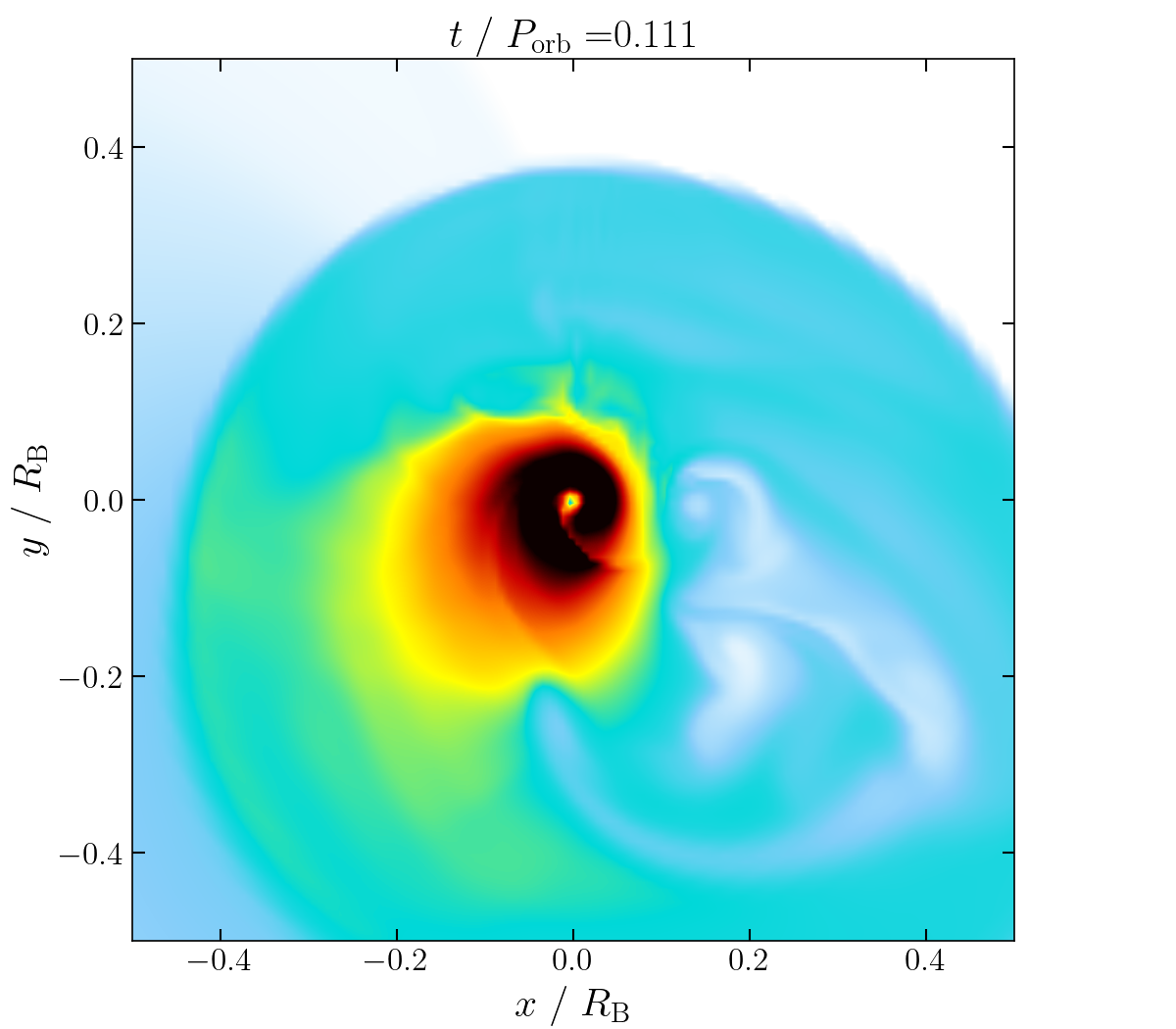}
		 	\includegraphics[width=0.65\columnwidth]{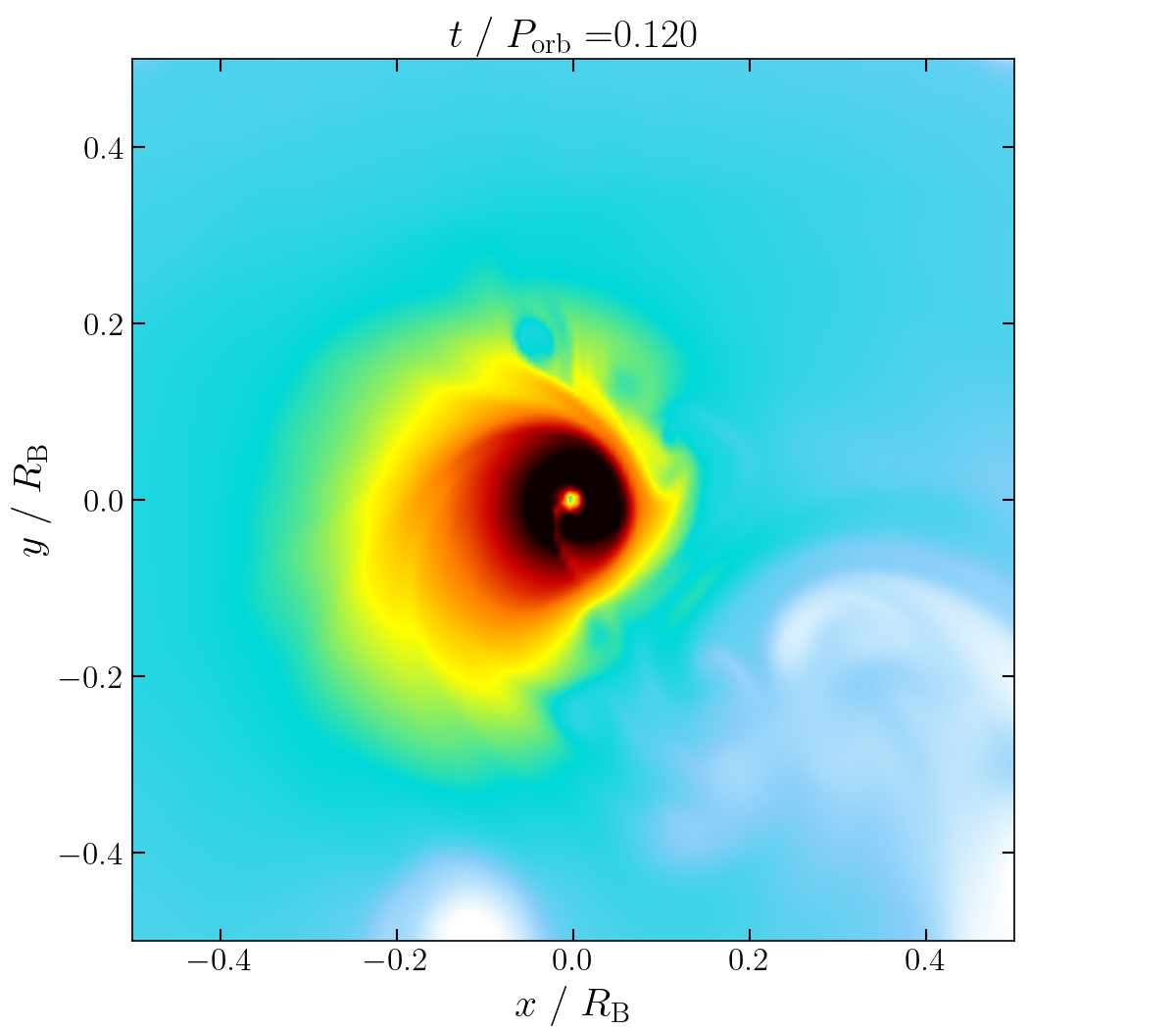}
		 	\includegraphics[width=0.65\columnwidth]{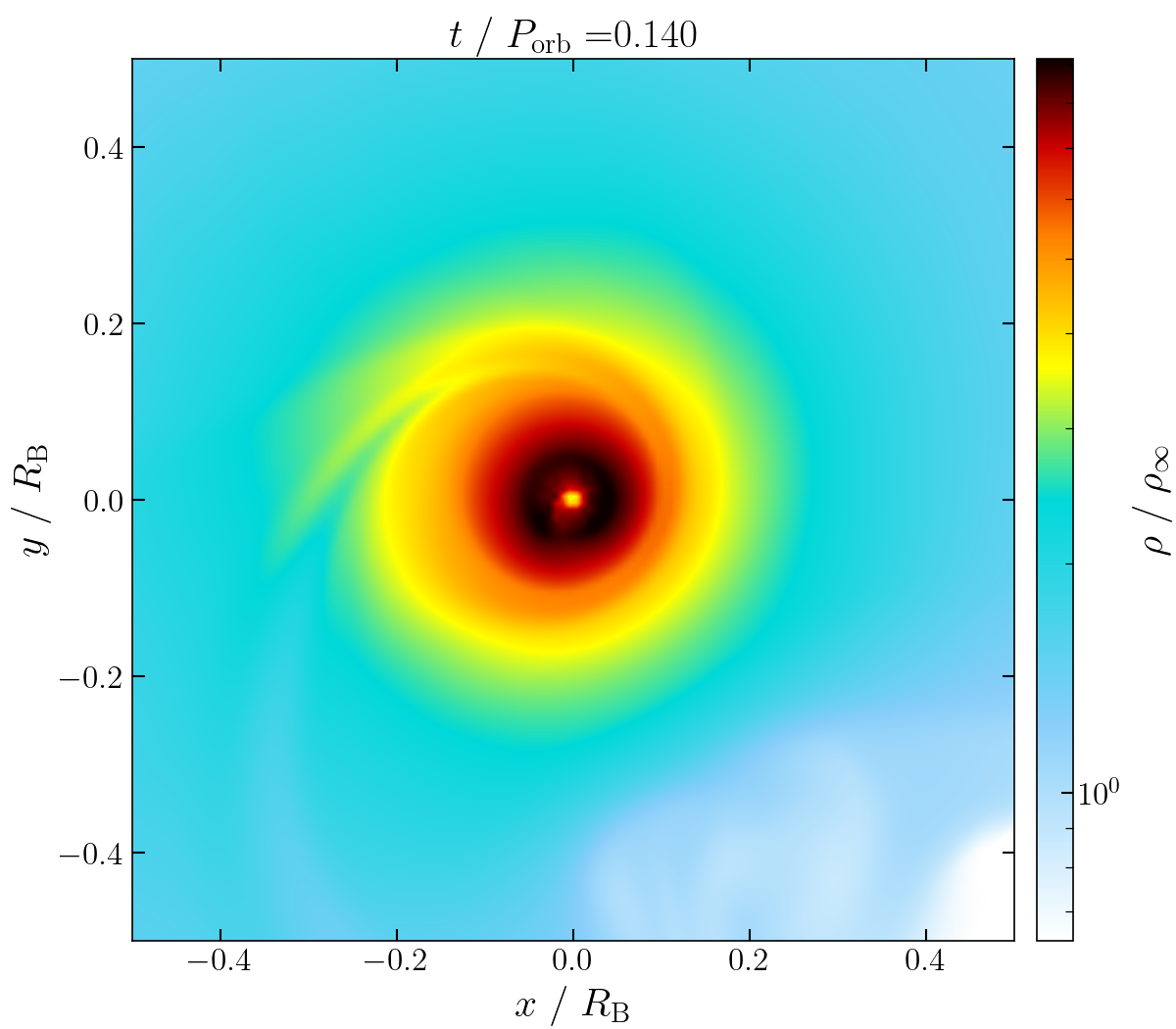}	
		 	\caption{The gas density contour at different times for a 2D simulation run with $m_{\bullet} = 50{\rm M}_{\odot}$, $\rho_{\rm cl} = 10^7{\rm M}_{\odot}/{\rm pc}^3$ and circular BH orbit. The BH orbital angular momentum vector is directed out of the page towards the reader. In the upper left panel, disk formation is not yet triggered. We observe a density front with a rotation tendency counter to the BH orbital rotation, generated by the transverse velocity shear induced by the BH rotation itself. At the time of the middle upper panel, disk formation is triggered and proceeds as a rapidly evolving angular momentum wave from inner to outer radii in the next panels. At the time of the bottom right panel, the steady state of a disk, counter-rotating to the BH orbit, is reached.}
		 	\label{fig:movie_2D}
		 \end{figure*}
		 
		 The angular momentum injection is subject to non-inertial forces within the Bondi sphere, which is advected along the BH trajectory as it is bound by the BH gravitational potential. The Coriolis force, Eq.~(\ref{eq:a_cor}), opposes angular momentum injection. On the BH orbital plane it is $a_{{\rm Cor},\phi} = -2\Omega v_r$ for a circular BH orbit. The timescale of the Coriolis effect is therefore $\tau_{\rm Cor} \sim R_{\rm B} / (2v_r)$ at $r\sim R_{\rm B}$. The disk-formation timescale cannot be shorter than $\tau_{\rm Cor}$ to ensure that a disk is persistent, $\tau_{\rm d} > \tau_{\rm Cor}$. Since the Coriolis timescale is one half the communication time, $R_{\rm B} / v_r$ -- from $R_{\rm B}$ down to the disk-formation radius $R_{\rm d}$ for $R_{\rm d} / R_{\rm B} \ll 1$ -- we can anticipate that the disk formation timescale is of order the crossing timescale, $\tau_{\rm cross} = R_{\rm B} / V_{\bullet}$. This is because at $r\sim R_{\rm B}$  the highest radial inward velocity is $V_{\bullet}$, therefore, we expect that
 		\begin{equation}
 			\tau_{\rm d} \sim \lambda_{\tau}
 			\tau_{\rm cross}\, ,\quad 
 			\tau_{\rm cross} = \frac{R_{\rm B}}{V_{\bullet}}.
 		\end{equation}
 		We denote $\lambda_{\tau}$ a prefactor of order one, to be estimated empirically.
 		
 		\begin{figure}[tbp]
 			\centering
 			\begin{subfigure}{\columnwidth}
 				\centering
 				\includegraphics[width=0.9\columnwidth]{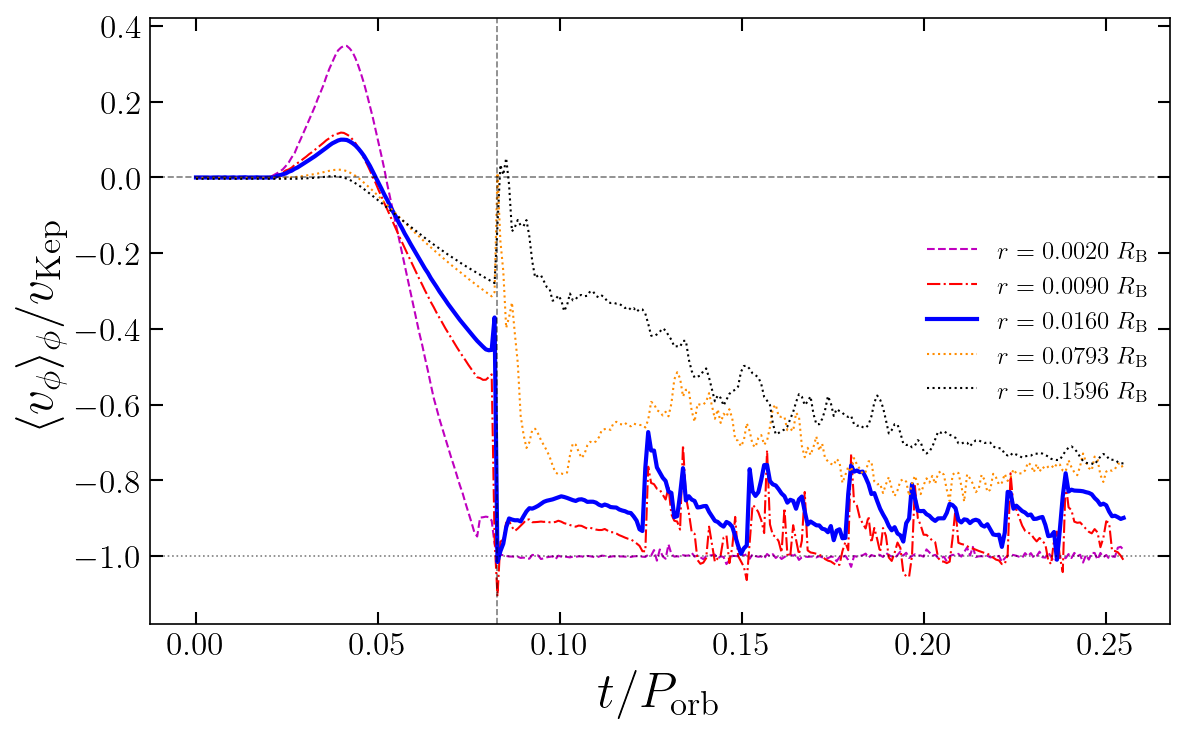}
 				\caption{$\psi_0 = 0^o$}
 				\label{fig:vphi_kep_overlay_ecc09_psi0_2D}
 			\end{subfigure}
 			\begin{subfigure}{\columnwidth}
 				\centering
 				\includegraphics[width=0.9\columnwidth]{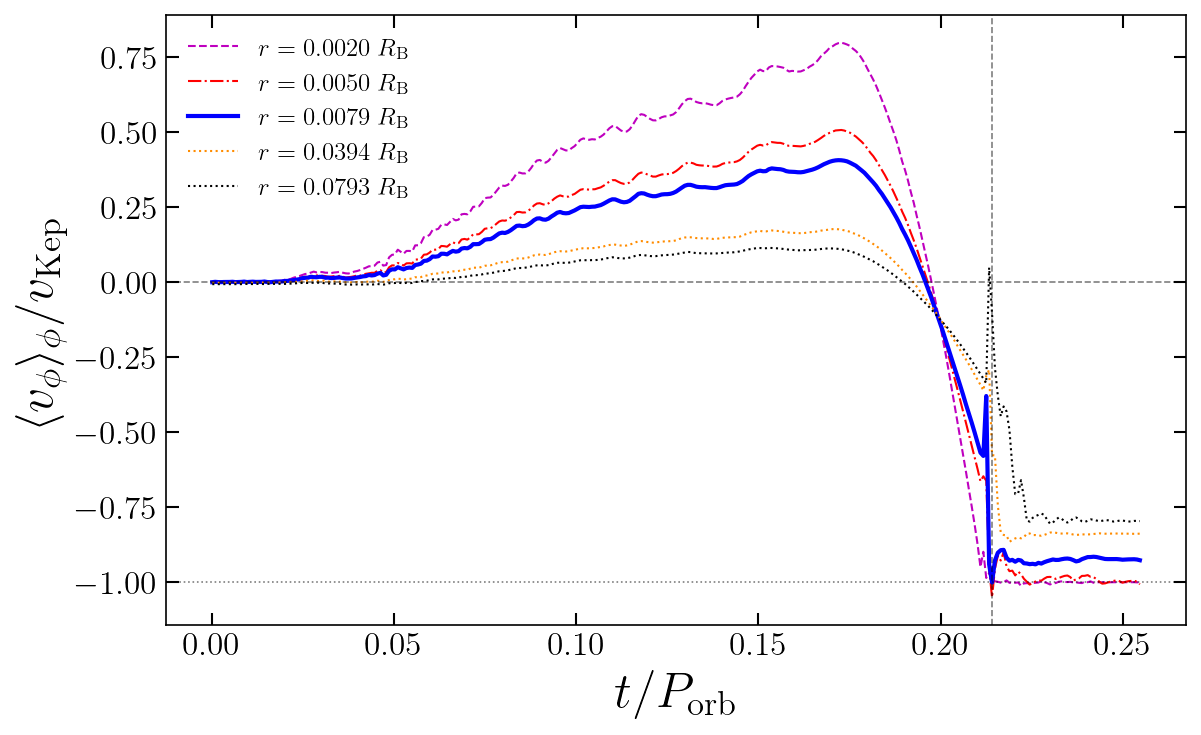}
 				\caption{$\psi_0 = 90^o$}
 				\label{fig:vphi_kep_overlay_ecc09_psipi2_2D}
 			\end{subfigure}
 			\caption{A 2D simulation run for a highly eccentric orbit $e = 0.9$ and $m_{\bullet} = 50{\rm M}_{\odot}$, $\rho_{\rm cl} = 10^7{\rm M}_{\odot}/{\rm pc}^3$. Notation is the same as in Figure \ref{fig:v_time_2D}. In the upper panel we depict the case of initial position at the semi-major axis, and in the bottom panel at the semi-minor axis. 
 				In the former case, the acceleration term, $\dot{\omega}_{\bullet} > 0$, enhances rotation of the gas counter to the BH orbital motion inducing an earlier disk formation time (vertical gray dotted line) and higher disk radius (blue solid curve) than in the latter case (bottom panel) where $\dot{\omega}_{\bullet} < 0$ enhances gas' co-rotation with the BH orbit.}
 			\label{fig:vphi_kep_overlay_ecc09_2D}
 		\end{figure}
 		
		 \section{Simulations}\label{sec:sims}
		 
		 \begin{figure}[tbp]
		 	\centering
		 	\begin{subfigure}{\columnwidth}
		 		\centering
		 		\includegraphics[width=0.9\columnwidth]{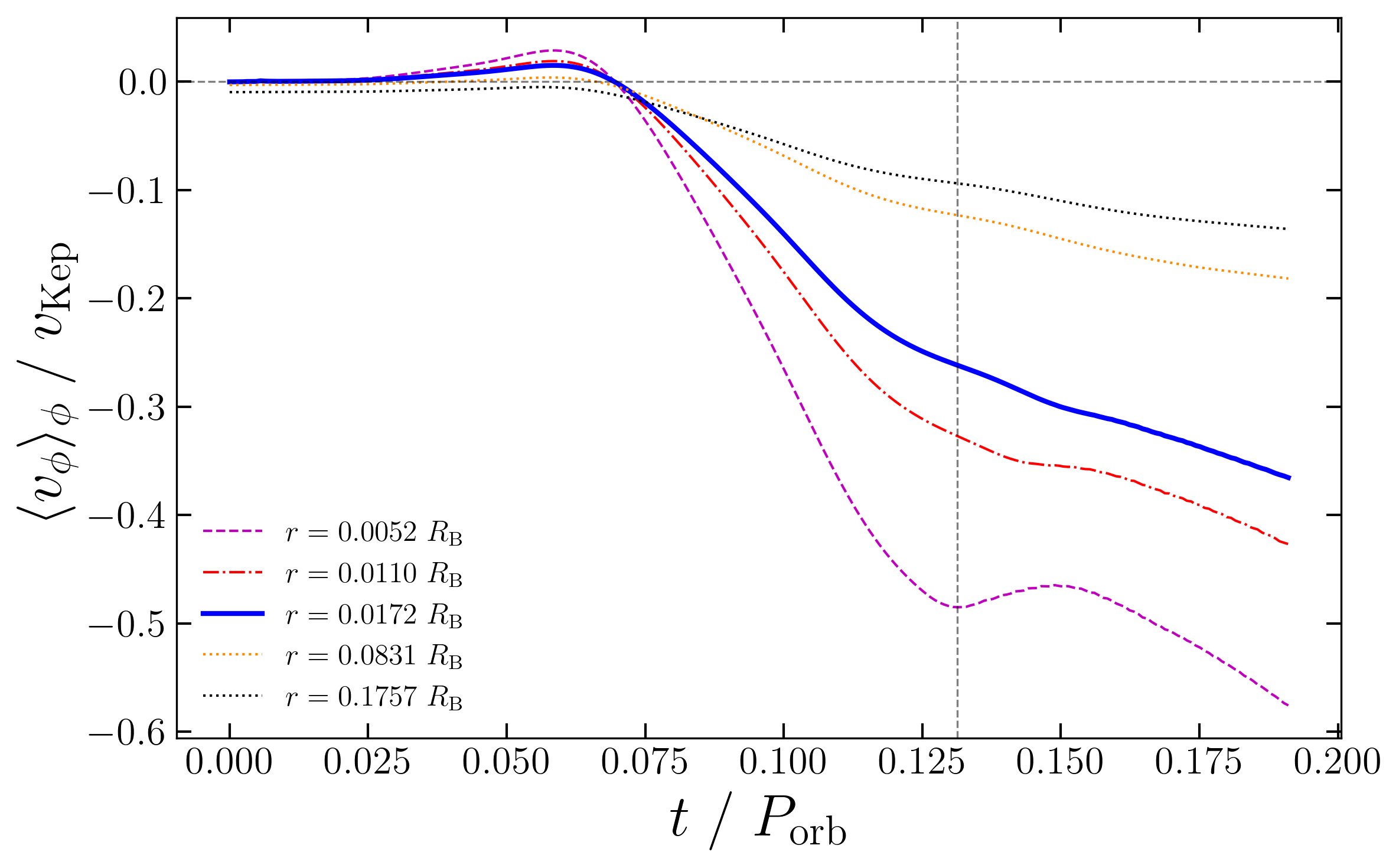}
		 		\caption{}
		 		\label{fig:vphi_time_3D}
		 	\end{subfigure}
		 	\begin{subfigure}{\columnwidth}
		 		\centering
		 		\includegraphics[width=0.9\columnwidth]{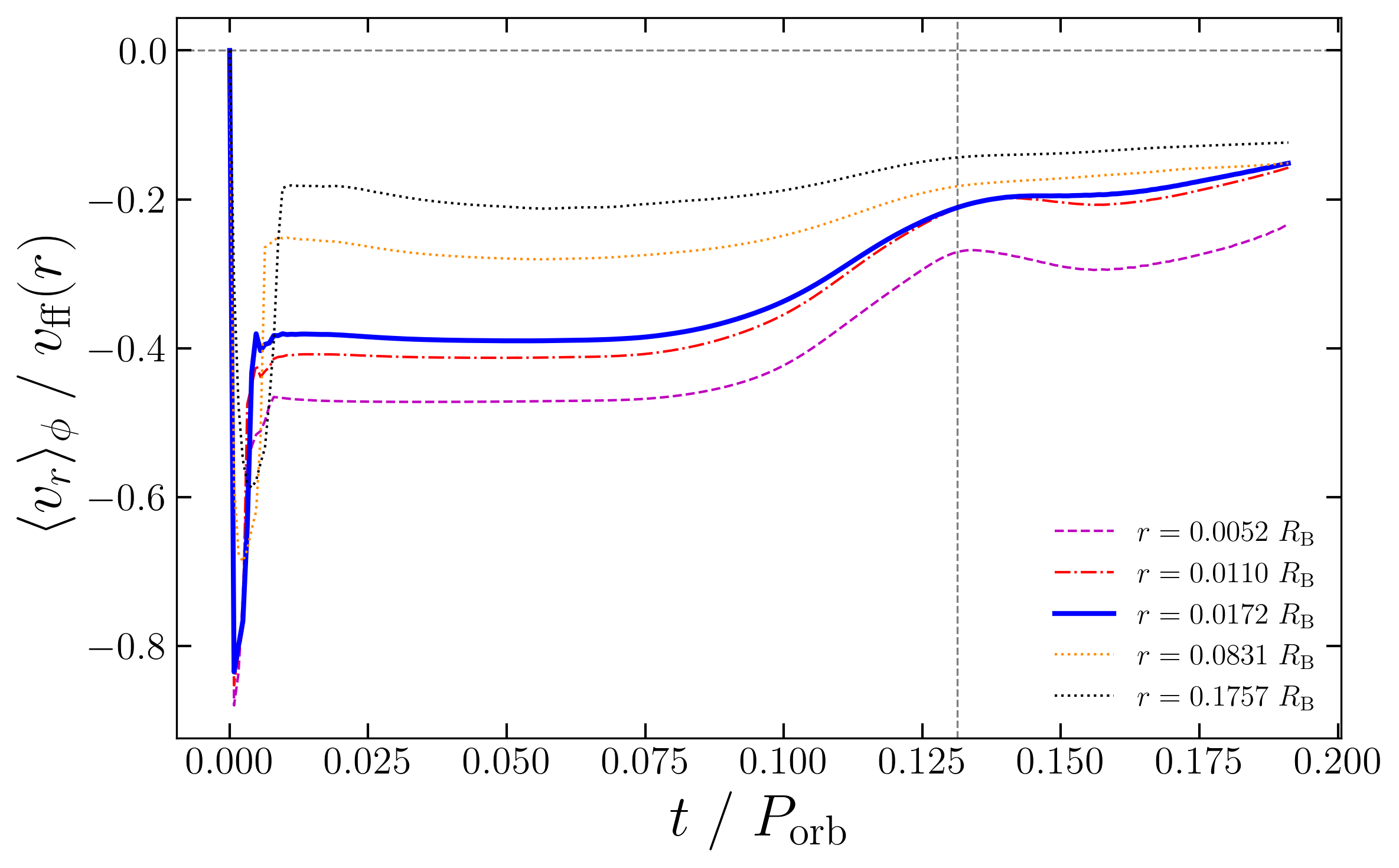}
		 		\caption{}
		 		\label{fig:vr_time_3D}
		 	\end{subfigure}
		 	\caption{A 3D simulation run for $m_{\bullet} = 50{\rm M}_{\odot}$, $\rho_{\rm cl} = 10^7{\rm M}_{\odot}/{\rm pc}^3$ and circular BH orbit. Notation is the same as in Figure \ref{fig:v_time_2D}. In the 3D case the disk formation trigger (vertical gray dotted line) is manifested as a bump in the inner radii of $\langle v_{\phi}\rangle_{\phi} / v_{\rm Kep}$ which occurs at $\sim -0.5$ in contrast to the 2D case where it occurred at the extreme value ($-1$). The blue solid curve designates evolution at the disk radius.}
		 	\label{fig:v_time_3D}
		 \end{figure}
		 
		  We performed 2D and 3D Newtonian hydrodynamic simulations using \textsc{Athena++} v21.0 \citep{2020ApJS..249....4S, athena_v21}\footnote{\url{https://github.com/PrincetonUniversity/athena}} with the goal of investigating accretion-disk formation. We developed a custom problem generator code, publicly available \citep{Roupas_athena_code_2026}\footnote{\url{https://github.com/zachariasroupas/protoBH-hydro}}, implementing the moving BH in its comoving reference frame $Bxyz$ within gas, which is subject to non-inertial forces (\ref{eq:a_r})-(\ref{eq:a_phi}). The outer boundary and initial conditions were given by Eq.~(\ref{eq:v_Sigma_an}). We used outflow inner boundary conditions. 
		  
		 The major challenge our simulations face in resolving the disk's outer radius during its formation is that the boundary conditions must lie outside the Bondi radius, while the disk itself forms deep within it. 
		 Let us estimate the disk radius heuristically.
		 Assuming the harmonic potential $\Phi(r) =(1/2) \Omega^2 r^2$, the BH probability density is $p_{\bullet}(r) \sim \exp \{-m_{\bullet} \Phi(r)) / \bar{m}_{\star} \sigma_{\star}^2$\}, where $\bar{m}_{\star}$ denotes the mean individual star mass and the velocity dispersion may be approximated by Virial theorem as $\sigma_{\star}^2 \sim \pi G M_{\star} / 32 r_{c}$. 
		 The root-mean-square (rms) BH position is then 
		 \begin{equation}
		 	R_{\bullet,{\rm rms}} = \sqrt{3\frac{\bar{m}_{\star}}{m_{\bullet}}}\frac{\sigma_{\star}}{\Omega} \approx 
		 	0.03\, 
		 	r_c\, 
		 	\left( \frac{\varepsilon}{0.3}\right)^{\frac{1}{2}}
		 	\left( \frac{\bar{m}_{\star}}{0.5}\right)^{\frac{1}{2}} 
		 	\left( \frac{m_{\bullet}}{50}\right)^{-\frac{1}{2}},
		 \end{equation}
		 where $\varepsilon$ denotes the star formation efficiency.
		 The BH Mach number, $\mathcal{M}_{\bullet} \equiv V_{\bullet} / c_s$, at the rms radius is then
		\begin{align}
			\bar{\mathcal{M}}_{\bullet} &= \sqrt{3 \frac{\bar{m}_{\star}}{m_{\bullet}}} \frac{\sigma_{\star}}{c_s}
			\nonumber \\
			&\approx 0.3	  
			\left( 
			\frac{\nicefrac{\bar{m}_{\star}}{m_{\bullet}}}{0.01}
			\right)^{\frac{1}{2}}
			\mu_{\rm eff}^{\frac{1}{2}}
			\left(
			\frac{\gamma}{\nicefrac{5}{3} }
			\right)^{-\frac{1}{2}}
			\left(
				\frac{T}{10^4{\rm K}}
			\right)^{-\frac{1}{2}}
			\left( 
			\frac{ M_{\star} / r_c }{10^6 {\rm M}_{\odot} /{\rm pc}}
			\right)^{\frac{1}{2}},
			\label{eq:Mach_BH}
		\end{align}
		where $\mu_{\rm eff}$ denotes the mean molecular weight and $\gamma$ is the adiabatic index. 
		Using Eq.~(\ref{eq:R_d}), the disk radius can therefore be estimated to be		 
		 \begin{align}
		 	\frac{R_{\rm d}}{R_{\rm B}} &\sim \lambda_R \frac{\Omega^2 R_{\rm B}^3}{G m_{\bullet}} 
		 	\nonumber \\
		 	&\approx 
		 	2\cdot 10^{-3}\,\lambda_R 
		 	\frac{\rho_{\rm cl}}{10^6\frac{{\rm M}_\odot}{{\rm pc}^3}}
		 	\left(\frac{m_{\bullet}}{50{\rm M}_{\odot}} \right)^2
		 	\left(
		 	\frac{ \gamma}{\nicefrac{5}{3}} 
		 	\right)^{-3}
		 	\left(
		 	\frac{T}{10^4{\rm K}}
		 	\right)^{-3}
		 	\left(\frac{\mu_{\rm eff}}
		 	{1 + \bar{\mathcal{M}}_{\bullet}^2}
		 	\right)^{3}
		 	\label{eq:R_d_det}
		 \end{align}
		 Throughout the following we assumed star formation efficiency $\varepsilon = 0.3$ and ambient gas temperature $10^4{\rm K}$ of an adiabatic ($\gamma = 5/3$), partially ionized ($\mu_{\rm eff} = 1$) H/He gas. 
		 
		 Therefore, this vast range of scales renders the simulations computationally demanding. The inner boundary has to be at most one fifth the estimated disk radius of Eq.~(\ref{eq:R_d_det}), and the outer boundary greater than $R_{\rm B}$ in order to capture disk formation.
		 We are motivated to perform not only 3D but also 2D simulations, which let us survey a wider range of density and BH mass values. The 2D runs also pin down the disk properties precisely, serving as invaluable comparative benchmarks for the 3D simulations. 

	Based on Eq.~(\ref{eq:R_d_det}), and as we explain below in more detail, we used values for the inner radius of our boundary conditions, $R_{\rm in} \sim 10^{-3}R_{\rm B}$, or larger. We get that the Schwarzschild radius $R_{\rm Schw}$ is many orders of magnitude smaller than our smallest probed scale, $R_{\rm in}$. We have $R_{\rm Schw} / R_{\rm in} \sim 10^3\, (c_s/c)^2 \sim 10^{-6}$ for our $c_s = 11.7\,{\rm km}/{\rm s}$. Therefore, the Newtonian approximation is perfectly valid. 
		 
		 \begin{figure*}[tbp]
		 	\centering
		 	\includegraphics[width=0.65\columnwidth]{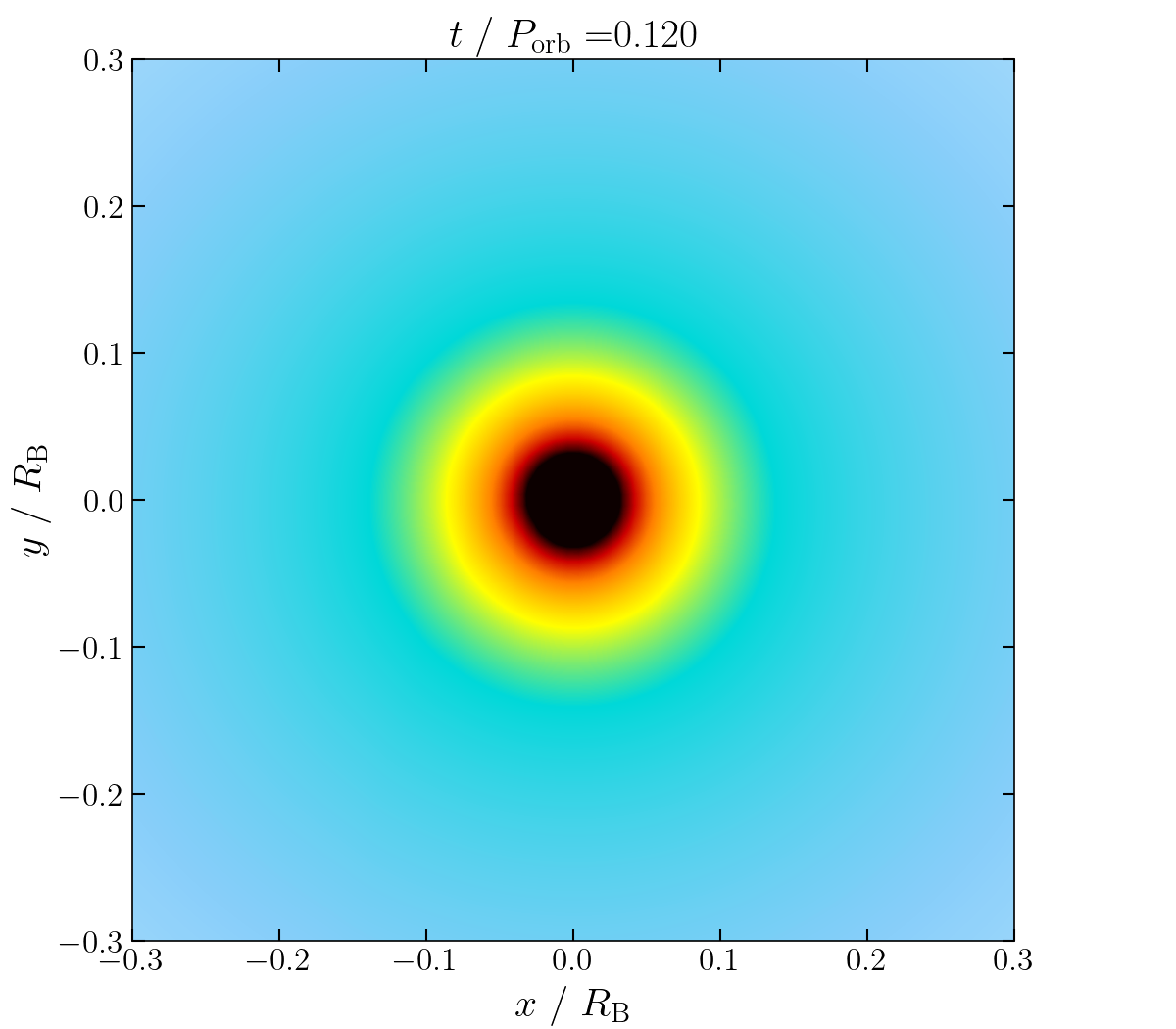}
		 	\includegraphics[width=0.65\columnwidth]{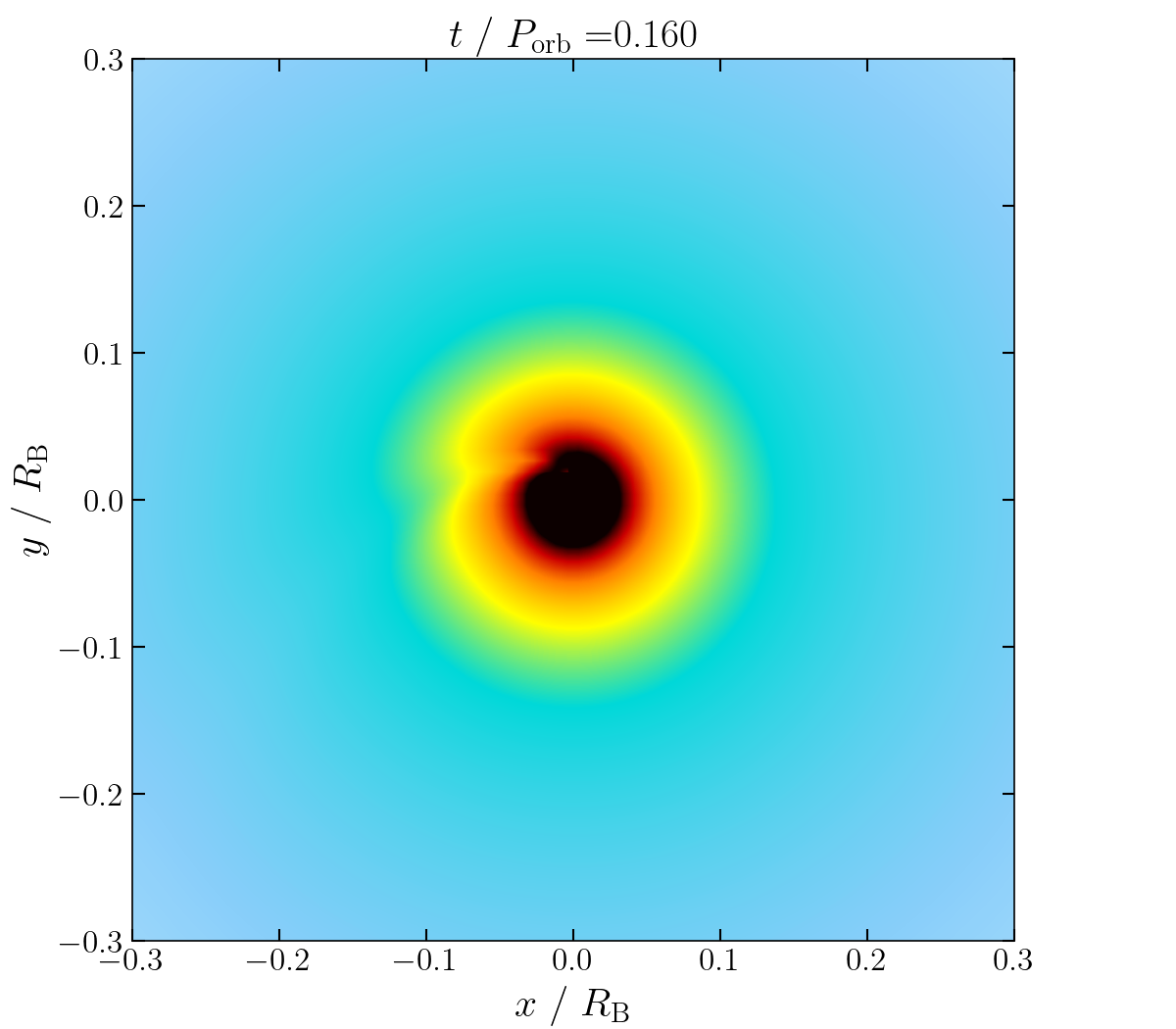}
		 	\includegraphics[width=0.65\columnwidth]{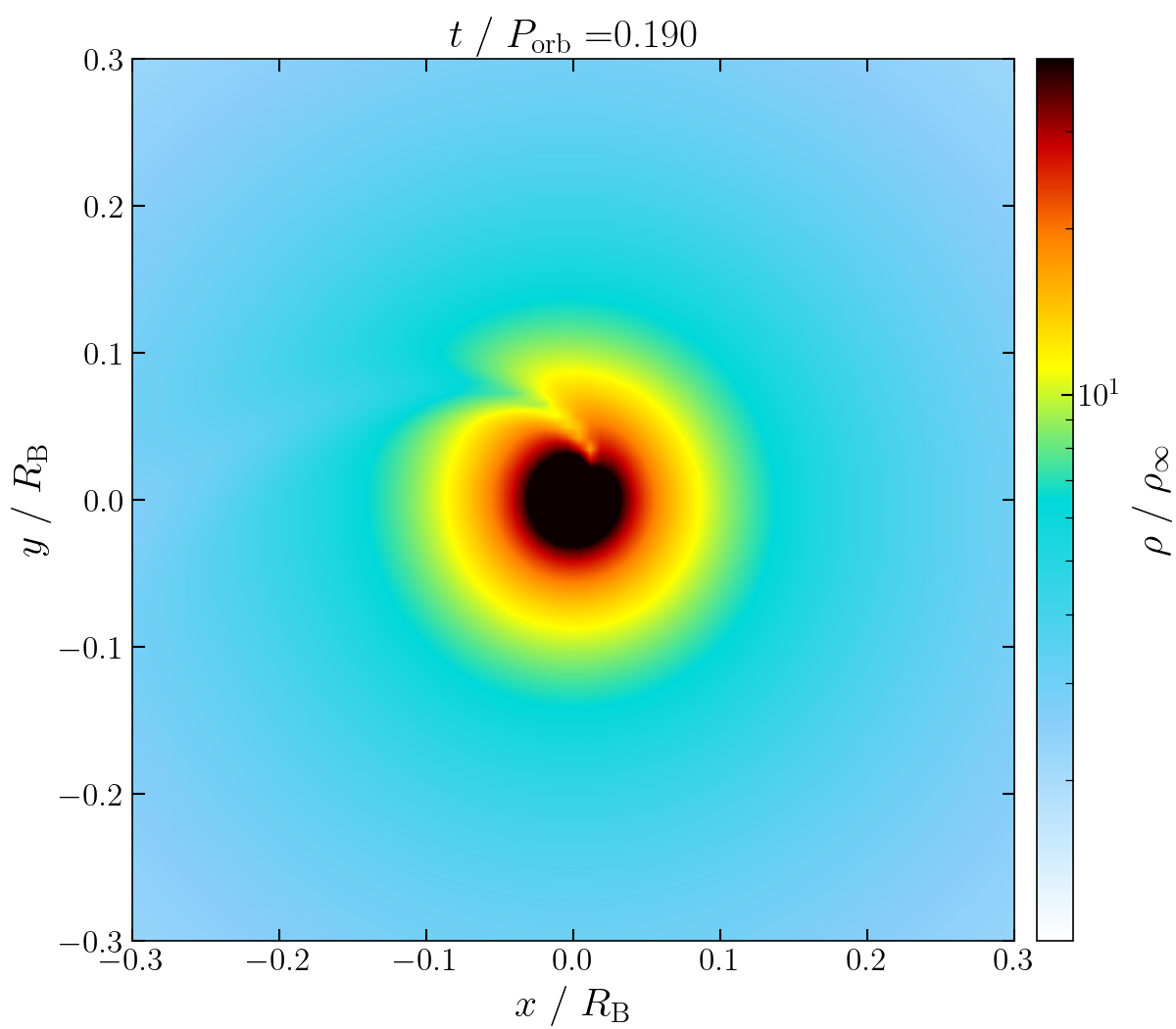}
		 	\caption{The gas density contour at the orbital plane at different times for a 3D simulation run with $m_{\bullet} = 50{\rm M}_{\odot}$, $\rho_{\rm cl} = 10^7{\rm M}_{\odot}/{\rm pc}^3$ and circular BH orbit. The BH orbital angular momentum vector is directed out of the page towards the reader. In contrast to the 2D case (Figure \ref{fig:movie_2D}), the disk formation trigger (medium panel) is not such a violent event, smoothed by $\theta$ contributions to the gas evolution. In the right panel the gas disk has formed, rotating counter to the BH orbit, approaching a steady state.}
		 	\label{fig:movie_3D}
		 \end{figure*}
		 
		 		  		\begin{figure}[tbp]
		 	\centering
		 	\begin{subfigure}{\columnwidth}
		 		\centering
		 		\includegraphics[width=0.9\columnwidth]{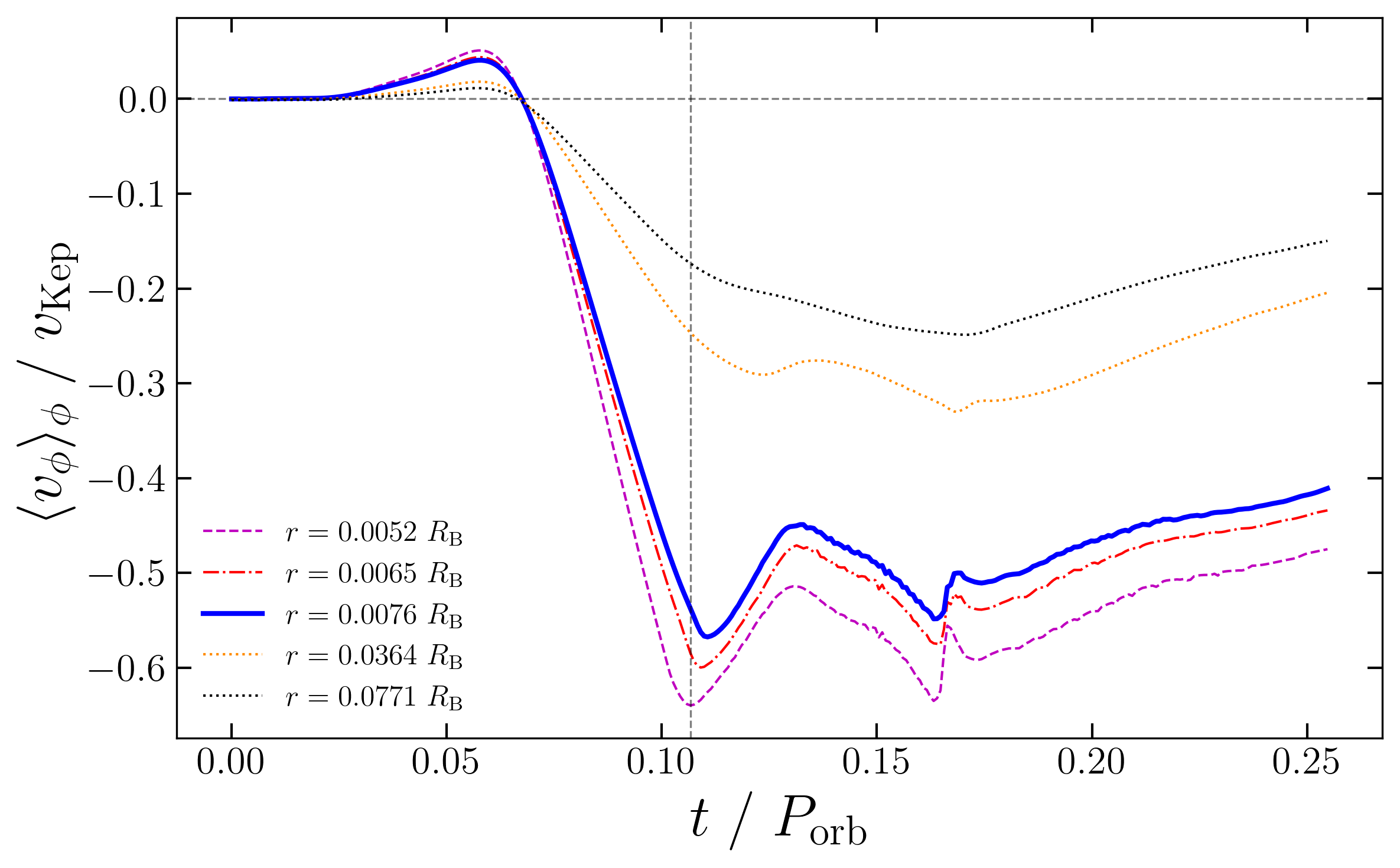}
		 		\caption{$\psi_0 = 0^o$}
		 		\label{fig:vphi_kep_overlay_ecc09_psi0_3D}
		 	\end{subfigure}
		 	\begin{subfigure}{\columnwidth}
		 		\centering
		 		\includegraphics[width=0.9\columnwidth]{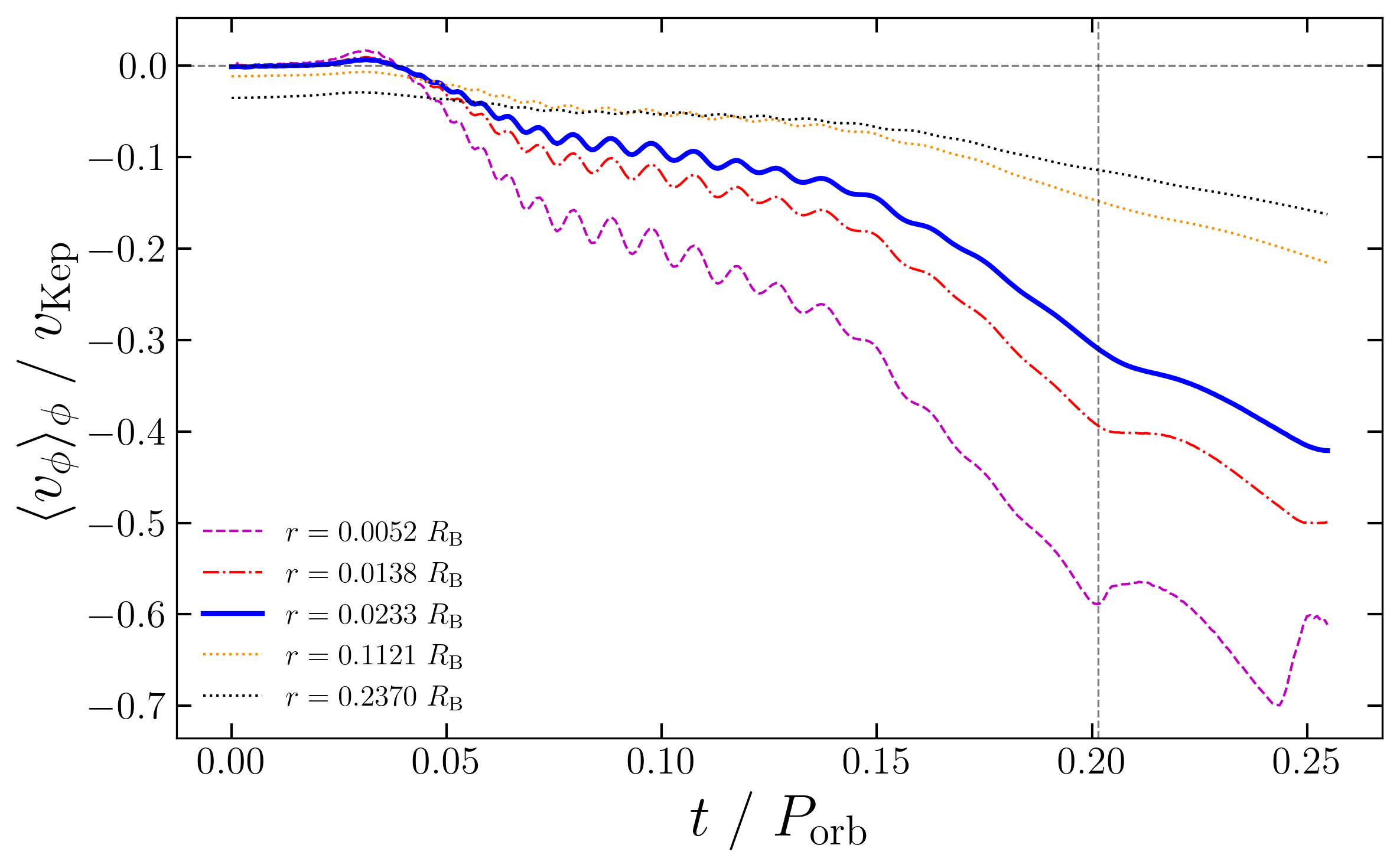}
		 		\caption{$\psi_0 = 90^o$}
		 		\label{fig:vphi_kep_overlay_ecc09_psipi2_3D}
		 	\end{subfigure}
		 	\caption{A 3D simulation run for a highly eccentric orbit $e = 0.9$ and $m_{\bullet} = 50{\rm M}_{\odot}$, $\rho_{\rm cl} = 10^7{\rm M}_{\odot}/{\rm pc}^3$. Similar to the 2D case, Figure \ref{fig:vphi_kep_overlay_ecc09_2D}. In the upper panel we depict the case of initial position at the semi-major axis, and in the bottom panel at the semi-minor axis. 
		 		In the former case, the acceleration term, $\dot{\omega}_{\bullet} > 0$, enhances rotation of the gas counter to the BH orbital motion while in the latter case (bottom panel), $\dot{\omega}_{\bullet} < 0$, counter-rotation is resisted.}
		 	\label{fig:vphi_kep_overlay_ecc09_3D}
		 \end{figure}
		 
		We want to set cluster density values and the BH position that are, on the one hand, physically realistic and, on the other, correspond to disk formation within at most three orders of magnitude of outer to inner boundary $R{\rm out}/R_{\rm in}$, so that our simulations remain computationally feasible. 
		The proto-stellar clusters observed by JWST in the Cosmic Gems arc galaxy at high redshift have typical star masses of $\sim 10^6{\rm M}_\odot$ and size $\sim 1{\rm pc}$, while they appear to be probably devoid of gas \citep{2024Natur.632..513A}. Therefore, the gaseous precursor should have mass $M_{\rm cl} \sim 3.3\cdot 10^6{\rm M}_\odot$ and should be more compact because the cluster expands during gas depletion \citep{2025A&A...702A.208R}. 
		 
	 In our 2D simulations, we considered the range of cluster total core density values $\rho_{\rm cl} = (5-10)\cdot 10^6{\rm M}_\odot /{\rm pc}^3$ --including both star and gas components-- , and set the BH position to $R_{\bullet} = 0.01 {\rm pc}$. 
	 The cluster density together with the BH position determine the BH orbital speed; therefore it is not always useful to vary them simultaneously. Those values restrict the BH speed to $\mathcal{M}_{\bullet} = 0.25-0.36$, consistent with our theoretical estimate, Eq.~(\ref{eq:Mach_BH}). 
	 We have additionally inspected the case of sonic BH speed $\mathcal{M}_{\bullet} = 1$ in 3D for the extreme case $\rho_{\rm cl} = 8\cdot 10^7{\rm M}_\odot /{\rm pc}^3$ with $R_{\bullet} = 0.01R_{\rm B}$, $m_{\bullet} = 100\,{\rm M}_{\odot}$. We verified that a disk still forms even at sonic speeds, without any evidence of a flip-flop instability \citep{1987MNRAS.226..785M,1988ApJ...335..862F,2005A&A...435..397F,2009ApJ...700...95B,2012ApJ...752...30B,2013ApJ...767..135B}.
	 There still remains a possibility that at significantly supersonic speeds a flip-flop instability may occur, which could  be a subject of future investigation. However, we stress that for our physical systems, supersonic speeds for regular BH orbits are unrealistic as we have already discussed.

	 The ambient gas density sets the boundary gas density $\rho_{\infty} = (1 - \varepsilon) \,\rho_{\rm cl}$ in our simulations.
	 We consider BH masses $m_{\bullet} = 15-100{\rm M}_{\bullet}$, intentionally covering values within the upper BH mass gap, $\sim 60-130{\rm M}_{\odot}$. This accounts for the possibility that stellar BHs have attained such values via accretion \citep{2025A&A...702A.208R}. We stress, however, that in this work we investigate solely angular momentum transfer and not mass accretion rate.

	 In the 2D simulations we used mostly an $r\times\phi$ grid of $168\times 304$, while in some cases (movies and special plots) we set a denser grid, $168\times 384$.
	 In the 3D simulations our denser  $r\times\theta\times\phi$ grid was $128\times 105\times 240$. Mostly, we used a grid,  $80\times 84\times 160$, and in demanding cases, where $R_{\rm d}/R_{\rm out} < 10^{-3}$ or that required running for a longer time, we used $80\times 63\times 128$. 
	 
	 We have tested several outer boundary radii in the range $1<R_{\rm out}/R_{\rm B} \leq 5$ and have found that our results are not sensitive to the boundary 
	 as long as $R_{\rm out} /R_{\rm B} \gtrsim 1.2$, 
	 given that $R_{\rm circ} \lesssim 0.1 R_{\rm B}$. 
	 We set $R_{\rm out}/R_{\rm B} = 2$, which is far enough from $R_{\rm B}$ but still not too large to make the simulations computationally inefficient. Our inspected range of density, BH position and BH masses serve well our physics motivation and allow us to inspect a feasible range of lengthscale $R_{\rm out}/R_{\rm in} \lesssim 5\cdot 10^3$, wherein the disk edge is located.

		 	\begin{figure}[tbp]
	\centering
	\includegraphics[width=0.9\columnwidth]{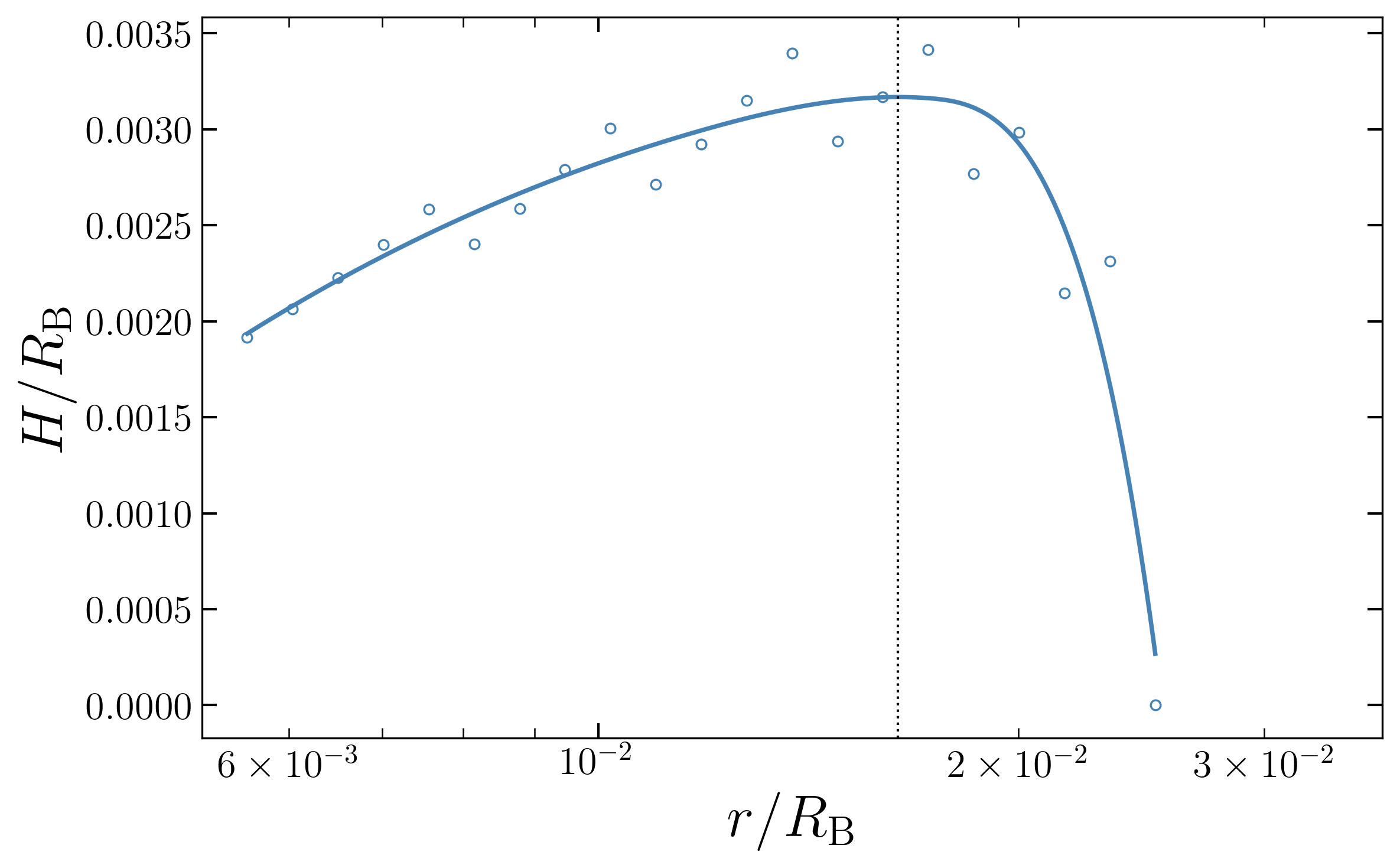}
	\caption{The marginal $z$-values we call disk height and denote $H$, which satisfy our disk-definition thresholds Eqs.~(\ref{eq:vr_threshold}), (\ref{eq:vphi_threshold}), with respect to radius. It is generated by a 3D simulation run with $m_{\bullet} = 50{\rm M}_{\odot}$, $\rho_{\rm cl} = 10^7{\rm M}_{\odot}/{\rm pc}^3$ and circular BH orbit, at the final simulation time $t_{\rm end} = 0.2P_{\rm orb}$. The data-points variability is an artifact of the limited resolution in the $\theta$ grid. The solid line is a smoothing fitting curve. The vertical gray dotted line represents the position of the maximum height. }
	\label{fig:H_r}
\end{figure}

\begin{figure}[tbp]
	\centering
	\includegraphics[width=0.9\columnwidth]{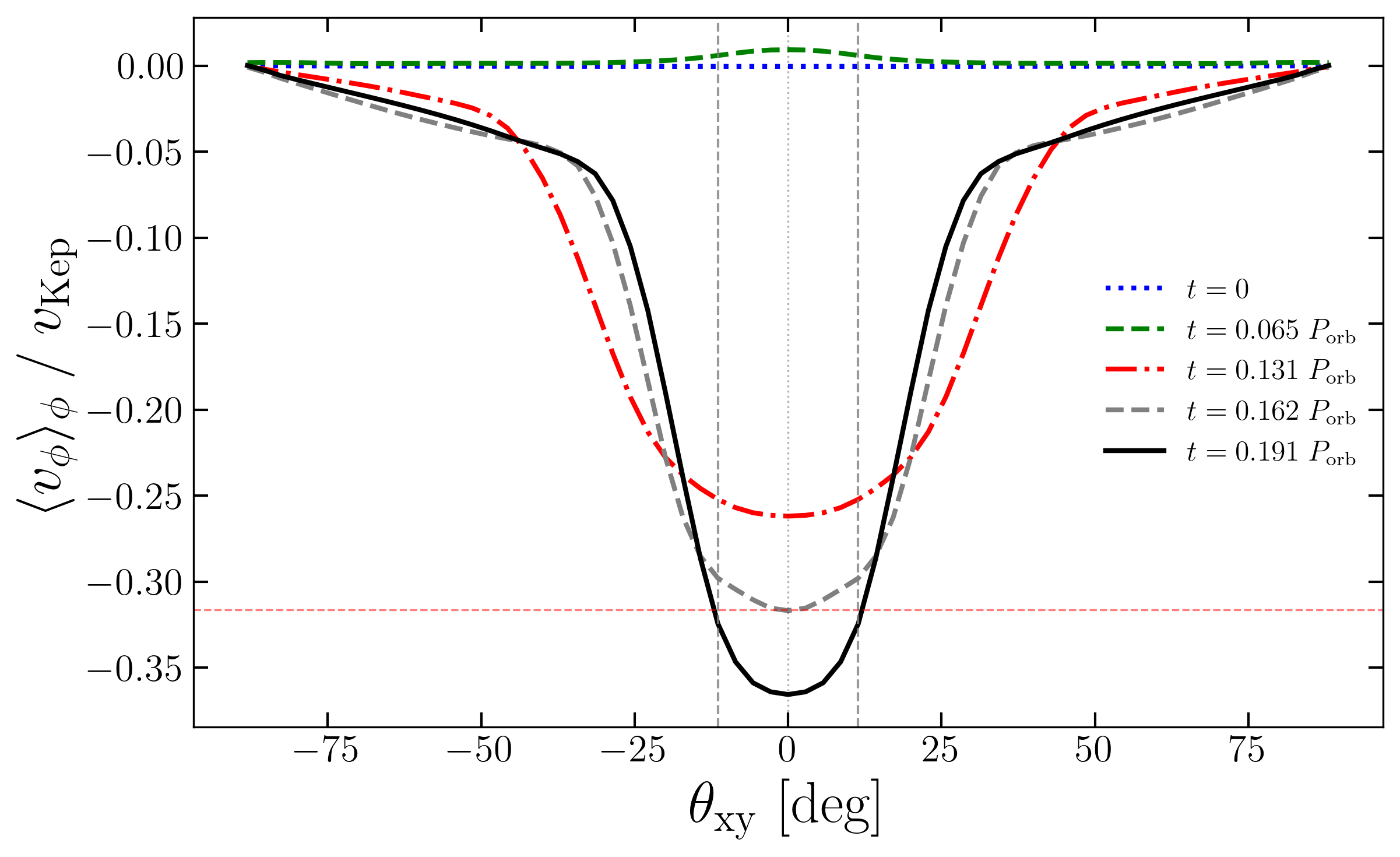}
	\caption{The average azimuthal velocity over the Keplerian velocity with respect to the opening angle for the estimated disk radius at several times. It is generated by a 3D simulation run for $m_{\bullet} = 50{\rm M}_{\odot}$, $\rho_{\rm cl} = 10^7{\rm M}_{\odot}/{\rm pc}^3$ and circular BH orbit. The vertical dashed line corresponds to the disk detection threshold.}
	\label{fig:vphi_theta_3D}
\end{figure}

	\subsection{2D}

A major advantage of the 2D simulations is that the disk radius and formation timescale can both be calculated unambiguously. In Figure \ref{fig:vphi_time_2D} we depict the azimuthal velocity average over $\phi$, $\langle v_{\phi} \rangle_{\phi}$, normalized by the Keplerian velocity $v_{\rm Kep}$ with respect to time at several radii for a circular orbit of a BH with mass $m_{\bullet} = 50{\rm M}_\odot$ and a cluster density $\rho_{\rm cl} = 10^7{\rm M}_{\odot} / {\rm pc}^3$. Remarkably, the disk formation time can be identified in a straightforward way by a sudden increase of $|\langle v_{\phi} \rangle_{\phi}|$ to values above $v_{\rm Kep}$ 
(below $-v_{\rm Kep}$ for $\langle v_{\phi} \rangle_{\phi}$) 
nearly simultaneously for a range of radii. The minus sign of $\langle v_{\phi} \rangle_{\phi}$ implies gas counter-rotation around the BH with respect to the orbital BH rotation. 
At the same time, the normalized radial velocity average over $\phi$ with respect to the free-fall time, $\langle v_{r} \rangle_{\phi} / v_{\rm ff}$, exceeds momentarily zero (Figure \ref{fig:vr_time_2D}) and oscillates afterwards around zero at those radii where gas becomes rotationally supported. 
Therefore, we define the disk formation timescale $\tau_{\rm d}$ and disk radius $R_{\rm d}$ directly as the earliest time when $\langle v_{\phi}(R_{\rm d}) \rangle_{\phi} / v_{\rm Kep}(R_{\rm d}) < 1$ and $\langle v_{r}(R_{\rm d}) \rangle_{\phi} / v_{\rm ff}(R_{\rm d}) > 0$, at the outermost radius $R_{\rm d}$ where these conditions hold. 

We infer that disk formation proceeds rapidly as an angular momentum wave of rotational support propagating from inner to outer radii. This is clearly evident in Figure \ref{fig:movie_2D} where the density contour is plotted at different times for a circular orbit. The cluster center lies in the negative $x$ direction and the BH is moving upwards in these panels (the BH orbital angular momentum vector points out of the page, toward the reader). In the upper left panel we can identify a density front rotating in the clockwise direction (counter to BH orbital rotation) induced by the BH orbital motion moments before the disk starts to form. The upper middle panel corresponds to the moment of ignition of disk formation. In the next panels the angular momentum wave propagates from inner to outer radii. In the last, bottom right panel, the disk has reached a steady state. 

Notice in Figure \ref{fig:vphi_time_2D} that initially $\langle v_{\phi} \rangle_{\phi}$ becomes positive (gas co-rotation with respect to BH orbital rotation) and only later gradually attains a negative sign. This is because of the Coriolis force, Eq.~(\ref{eq:a_cor}), which opposes gas rotation counter to the BH orbital motion. For an elliptical orbit, there is additionally the angular acceleration term, $a_{\rm acc} = -\dot{\omega}_{\bullet} r \sin\theta$ (Eq.~(\ref{eq:a_Sigma})), which periodically enhances or opposes counter-rotation during parts of the orbit when $\dot{\omega}_{\bullet} > 0$ or $\dot{\omega}_{\bullet} < 0$, respectively. For a harmonic orbit rotation occurs around the center of the ellipse, not its focus. Therefore, the orbital velocity and acceleration values are repeated every $P_{\rm orb}/2$. We expect that for a highly eccentric orbit the disk formation time and radius will depend on the initial position of the BH. In our inertial coordinate system $OXYZ$ (see Figure \ref{fig:coord}) the polar angle $\psi=0^o$ corresponds to the semimajor axis $X = +A$, $Y=0$. In Figure \ref{fig:vphi_kep_overlay_ecc09_psi0_2D} we depict the azimuthal velocity for a highly eccentric orbit, $e=0.9$, with initial position $\psi_0 = 0^o$, that is the BH initially at the semi-major axis, and in Figure \ref{fig:vphi_kep_overlay_ecc09_psipi2_2D} with initial position $\psi_0 = 90^o$, that is the BH initially at the semi-minor axis. We evolve the system for time $P_{\rm orb}/4$. In the former case ($\psi_0 = 0^o$) it is $\dot{\omega}_{\bullet} > 0$ $\Rightarrow$ $a_{\rm acc} < 0$; therefore the angular acceleration $\dot{\omega}_{\bullet}$ enhances counter-rotation which is manifested as a more rapid disk formation and larger disk radius than in the case $\psi_0 = 90^o$, where $\dot{\omega}_{\bullet} < 0$ $\Rightarrow$ $a_{\rm acc} > 0$. In this latter case, the angular deceleration enhances co-rotation, delaying disk formation, as seen in Figure  \ref{fig:vphi_kep_overlay_ecc09_psipi2_2D}. The result $R_{\rm d} \sim R_{\rm circ}$ still holds approximately for highly eccentric orbits as well, where now the circularization radius corresponds approximately to the position where disk formation is triggered. 

 		\begin{figure}[tbp]
	\centering
	\begin{subfigure}{\columnwidth}
		\centering
		\includegraphics[width=0.9\columnwidth]{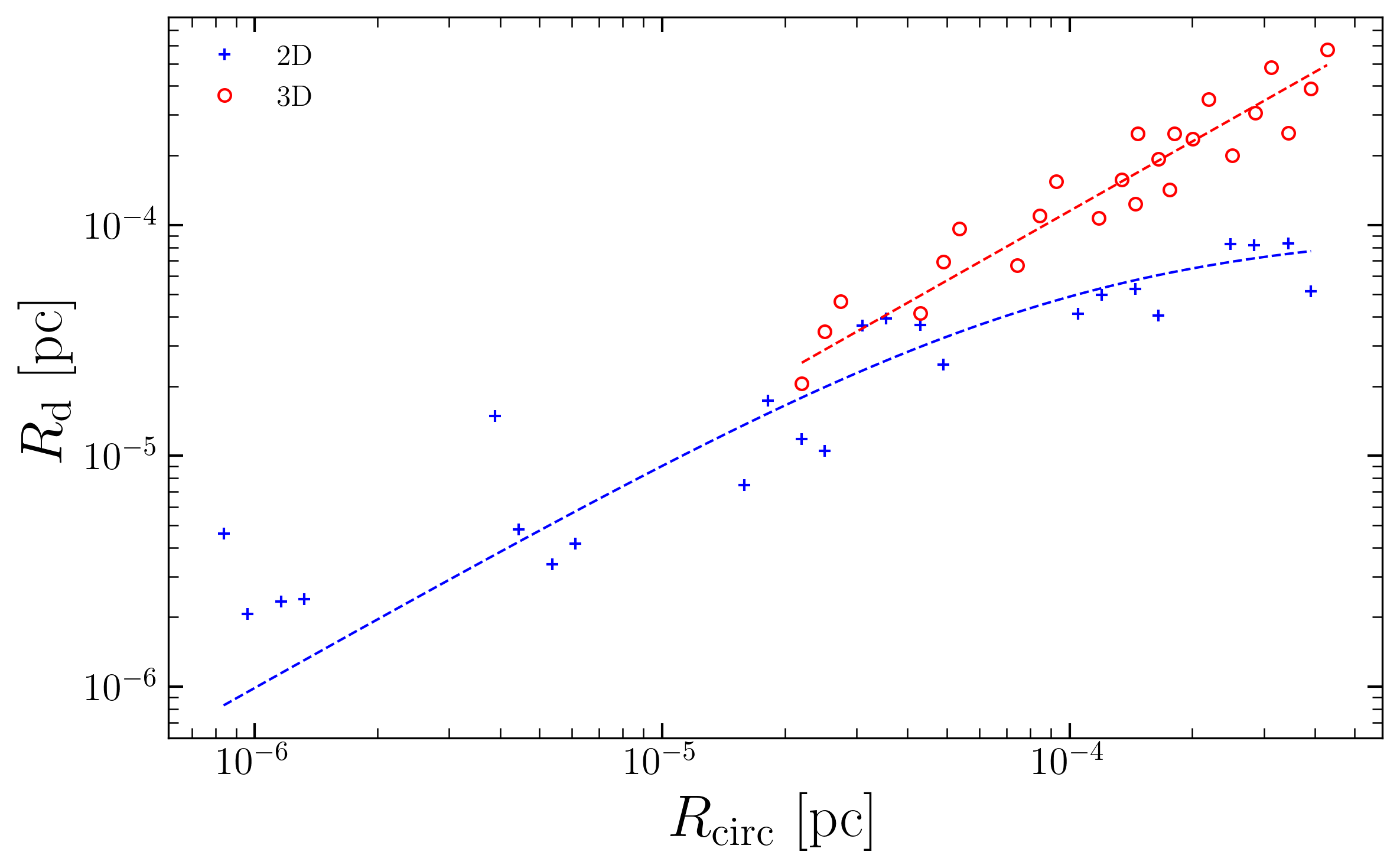}
		\caption{}
		\label{fig:Rd_Rc_pc}
	\end{subfigure} 
	\begin{subfigure}{\columnwidth}
		\centering
		\includegraphics[width=0.9\columnwidth]{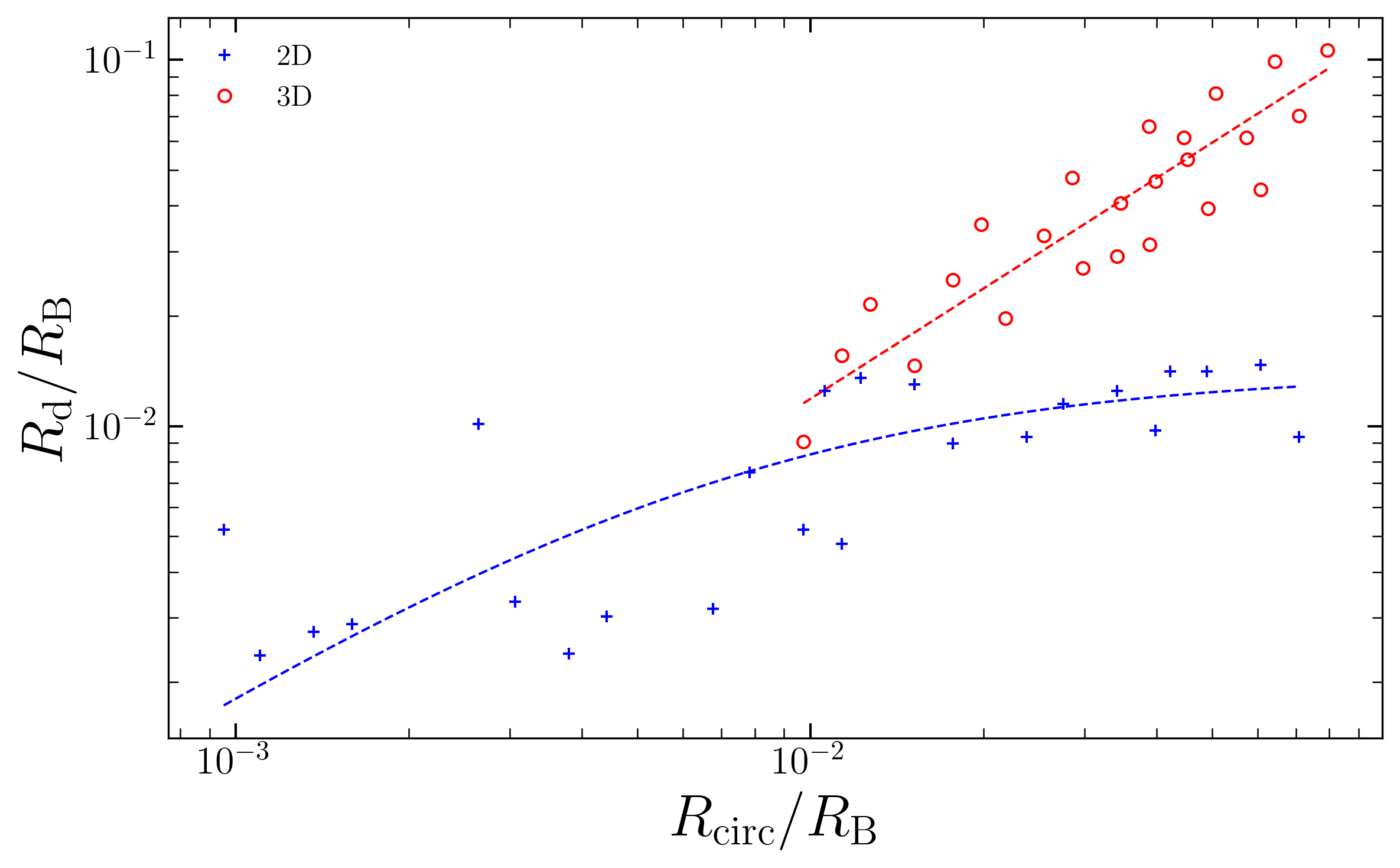}
		\caption{}
		\label{fig:Rd_Rc_Bondi}
	\end{subfigure}
	\caption{The estimated disk radius $R_{\rm d}$ inferred from our 2D and 3D simulations with respect to the theoretical circularization radius. The blue dashed line corresponds to a linear saturation fit in the 2D case, and the red dashed line to a linear fit in the 3D case. Upper panel in pc. Lower panel normalized by the Bondi radius.	}
	\label{fig:Rd_Rc}
\end{figure}

\begin{figure}[tbp]
	\centering
	\includegraphics[width=0.9\columnwidth]{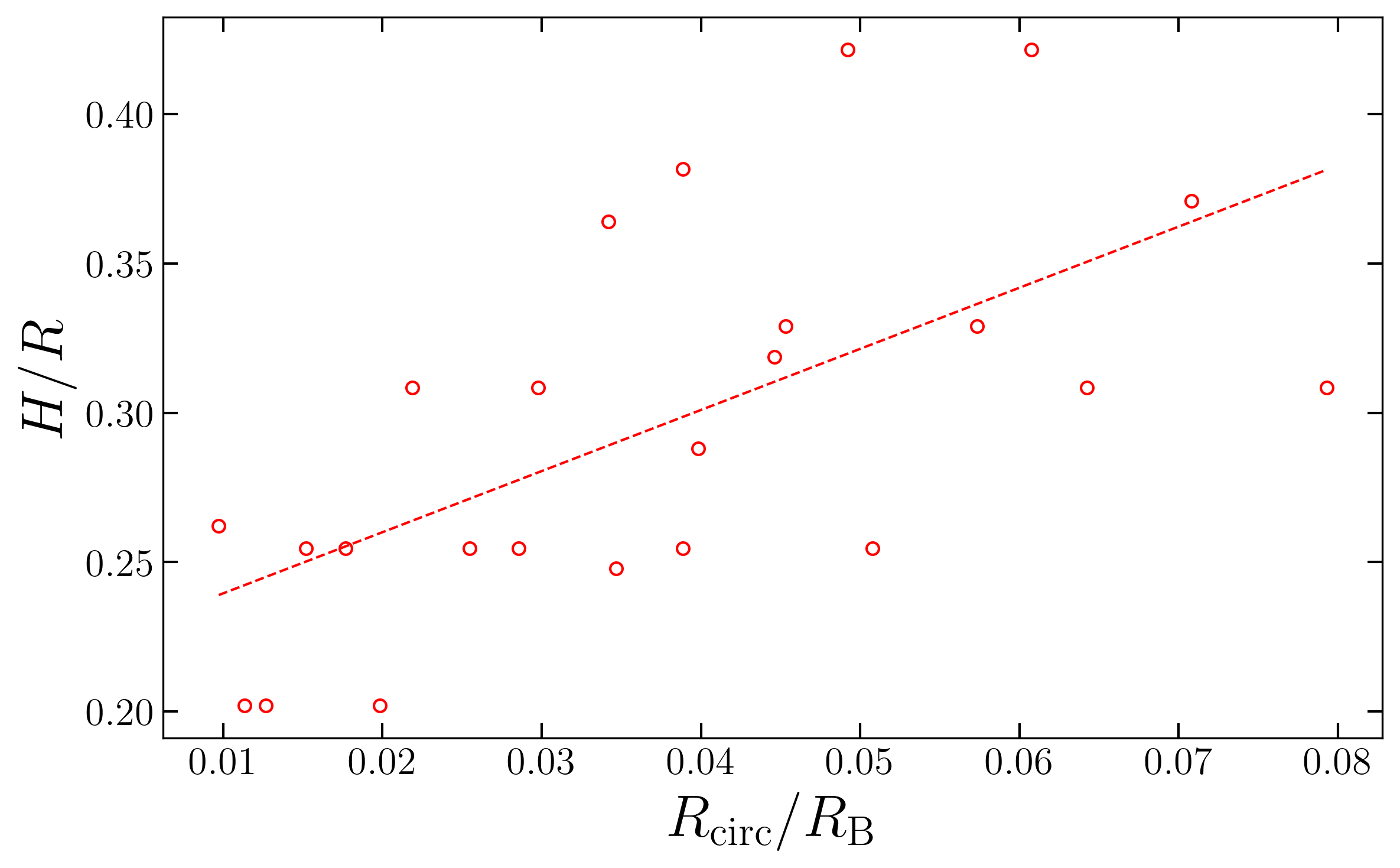}
	\caption{The disk aspect ratio with respect to the circularization radius normalized by the Bondi radius,  suggesting slim to thick disks. }
	\label{fig:HR}
\end{figure}

\begin{figure}[tbp]
	\centering
	\includegraphics[width=0.9\columnwidth]{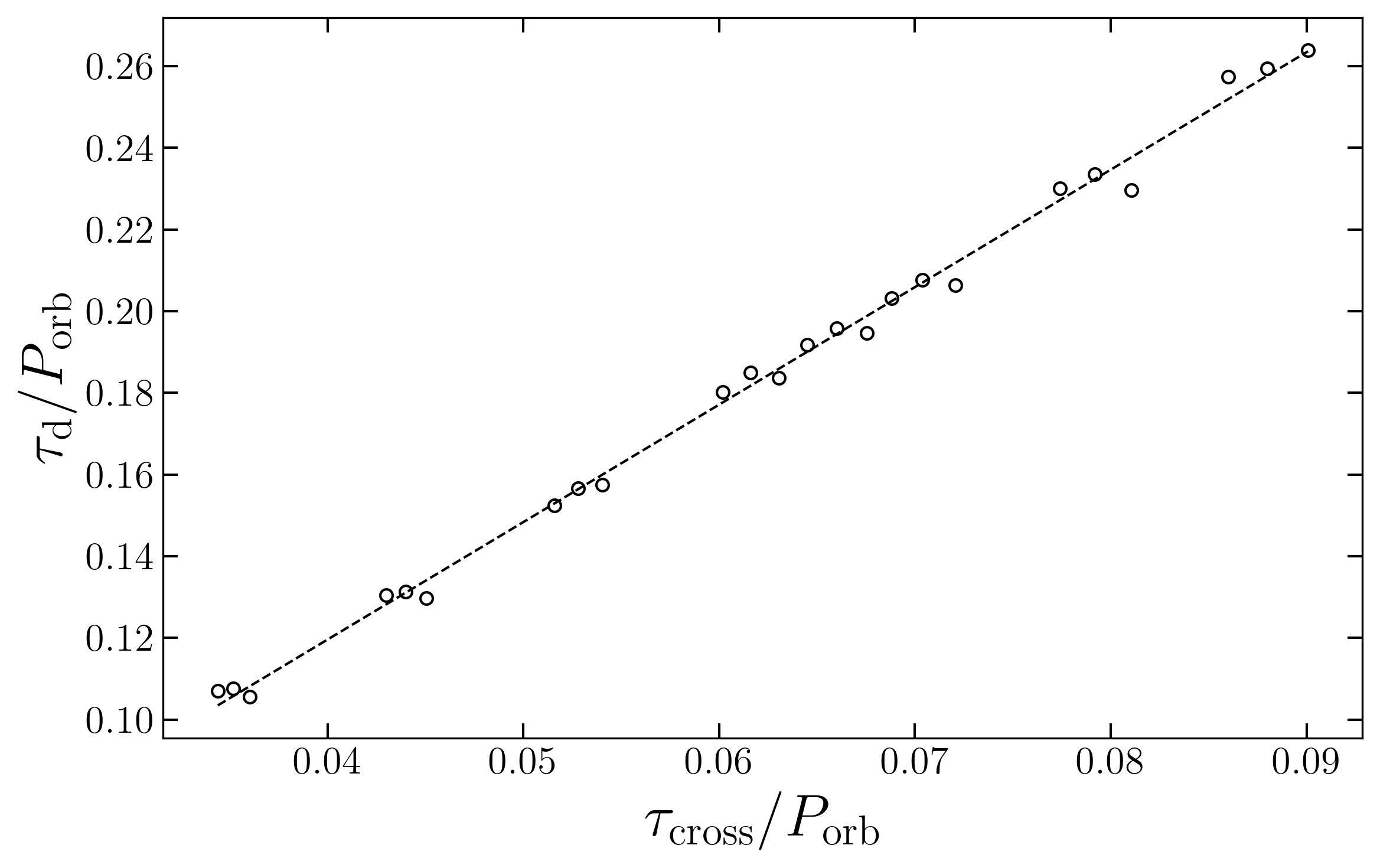}
	\caption{The estimated disk-formation ignition time $\tau_{\rm d}$ inferred from our 3D simulations with respect to the crossing timescale, $\tau_{\rm cross}$. }
	\label{fig:tau_d_vs_tau_cross_3D}
\end{figure}

		 \subsection{3D}
		 
		 In 3D simulations, disk formation is not ignited by a violent increase of the azimuthal velocity magnitude $|\langle v_{\phi} \rangle_{\phi} |$ to the value of the Keplerian velocity. Instead, it increases only progressively ($\langle v_{\phi} \rangle_{\phi}$ itself is negative like in 2D, manifesting disk counter-rotation to BH orbit) bouncing at some value of $\sim 0.5 v_{\rm Kep}$ in the innermost radii. This is evident in Figure \ref{fig:vphi_time_3D}. Compare with the 2D case, Figure \ref{fig:vphi_time_2D}, where disk formation is manifested as a violent variation of $\langle v_{\phi} \rangle_{\phi}$ to the value $-v_{\rm Kep}$. The smooth, gradual disk formation in 3D is also evident in Figure \ref{fig:movie_3D}, where we depict the density contour on the $xy$-plane at several times. The cause of this difference with respect to the 2D case is the contributions from the several $z$-planes.
		 
		 In Figure \ref{fig:vphi_kep_overlay_ecc09_3D} we depict the azimuthal velocity for a highly eccentric orbit $e=0.9$ and two different initial positions; at the semi-major and the semi-minor axis. Just like in the 2D case (Figure \ref{fig:vphi_kep_overlay_ecc09_2D}), gas counter-rotation to the BH orbit is enhanced when the BH is accelerating (moving from the semi-major towards the semi-minor axis) and resisted when it is decelerating (moving from the semi-minor towards the semi-major axis). The outer boundary of the accretion disk is less distinct in highly eccentric orbits than in a circular orbit, as $\langle v_{\phi} \rangle_{\phi}$ oscillates during a complete orbit.
		 
		 We identify the earlier time of a bounce as the disk formation time in the 3D case, $\tau_{\rm d}$. It is the clearest signal of disk formation ignition. Then, as the system evolves, the disk is slowly expanding outwards. The bouncing time is not sensitive to the choice of inner radius for the $R_{\rm in}$ scale used in our simulations. This is evident in Figure \ref{fig:v_time_3D} where we can see that $\langle v_{\phi}\rangle_{\phi}$ and $\langle v_{r}\rangle_{\phi}$ are flattened at different radii almost simultaneously. The identification of the disk radius is even less straightforward. We use a combination of selection criteria. Firstly, we require that the rotational energy is greater than the translational energy, $\langle v_{r}\rangle ^2 < \langle v_{\phi}\rangle_{\phi}^2$, which in our normalizations reads as
		 \begin{equation}\label{eq:vr_threshold}
		 	\frac{\langle v_{r}\rangle_{\phi}}{v_{\rm ff}}  <  \frac{1}{\sqrt{2}}\left| \frac{\langle v_{\phi}\rangle_{\phi}}{v_{\rm Kep}}\right|.
		 \end{equation}
		 In addition, we impose 
		 \begin{equation}\label{eq:vphi_threshold}
		 	\left(\frac{\langle v_{\phi}\rangle_{\phi}}{v_{\rm Kep}}\right)^2 > \beta_{\rm min}.
		 \end{equation}
		 We set $\beta_{\rm min} = 0.1$ to ensure that the rotational energy is of the same order of magnitude as the corresponding energy at the same $r$ of a Kepler orbit. This assignment corresponds to $R_{\rm d}$, such that $\langle v_{\phi}(R_{\rm d})\rangle_{\phi} /v_{\rm Kep}(R_{\rm d}) = - 0.31 $, $\langle v_r(R_{\rm d})\rangle_{\phi} / v_{\rm ff}(R_{\rm d}) = - 0.22$. We furthermore impose the constraint that at the final simulation time there exists a maximum of the height $H(r)$, defined as the marginal $z$ values which satisfy those thresholds. This $H$ profile is depicted in Figure \ref{fig:H_r}. We verify that $H(r)$ drops rapidly at $r > r_{\rm max}$, 
		 so that $\partial H/\partial r|_{R_{\rm d}} \rightarrow -\infty$.
		 In Figure \ref{fig:vphi_theta_3D} we depict the azimuthal velocity with respect to the opening angle $\theta_{\rm xy}$ of $H(R_{\rm d})$ with respect to the $xy$-plane at several times. We get similar results if we use the radius of the maximum height as the $R_{\rm d}$ definition. Therefore, our analysis is not sensitive to the precise value of $\beta_{\rm min}$ provided it is below the value of the disk-formation ignition bounce.
		 
		 In Figure \ref{fig:Rd_Rc} we depict the disk radius of several simulations for a circular orbit and both 2D and 3D cases. In 2D we have used the values $m_{\bullet} = 15, 25, 40, 50, 75, 100\,{\rm M}_\odot$, $\rho_{\rm cl} = 5,6,8,10\cdot 10^6{\rm M}_{\odot}/{\rm pc}^3$. 
		 The correlation coefficient of variables $R_{\rm d}$ and $R_{\rm circ}$ in the 2D case is $C_{\rm 2D} = 0.86$, suggesting a strong correlation. Nevertheless, $R_{\rm d}(R_{\rm circ})$ is better fit by a saturating exponential rather than by a linear function, tending to a linear dependence only for small circularization radius, 
		 $
		 	R_{{\rm d},{\rm 2D}} = \lambda_{R,{\rm 2D}} R_{\rm circ} /(1+ \nicefrac{R_{\rm circ}}{R_{\rm sat}})
		 $.
		 We get $\lambda_{R,{\rm 2D}} = 1.0$.
		 The cause of saturation is the Coriolis force (\ref{eq:a_cor}) which is perfectly effective in the 2D case at large radii close to $R_{\rm B}$. This is not the case in 3D, because Coriolis effect is weakened by several $\theta$ contributions. In 3D, we used the values  $m_{\bullet} = 40, 50, 60, 70, 75, 80, 90, 100\,{\rm M}_\odot$, $\rho_{\rm cl} = 0.8, 1.0, 1.2 \cdot 10^7{\rm M}_{\odot}/{\rm pc}^3$. 
		 The correlation coefficient of variables $R_{\rm d}$ and $R_{\rm circ}$ is $C_{\rm 3D} = 0.91$, suggesting a significant linear correlation. We get 
		 \begin{equation} 
		 	R_{\rm d} = \lambda_{R} R_{\rm circ}, \quad \lambda_{R} = 1.15
		 \end{equation}
		 for a linear fit through the origin. The spread of data points in Figure \ref{fig:Rd_Rc} is  caused by the different density values. We suspect this is an artifact of our limited simulation time. A higher gas density implies a shorter orbital period and a faster transfer of angular momentum from inner to outer radii. Following the bounce, which signifies disk formation, angular momentum is slowly propagating to outer radii (see Figure \ref{fig:vphi_time_3D}), expanding the radii for which our thresholds are satisfied. This process occurs with a different speed for a different density value.
		 
		 In Figure \ref{fig:HR}, we depict the aspect ratio $H/R = \tan(\theta_{{\rm xy},{\rm d}})$ with respect to $R_{\rm circ} / R_{\rm B}$. We denote $\theta_{{\rm xy},{\rm d}}$ the disk opening angle at the $v_{\phi}$ threshold and at the final time of the simulation. A higher density tends to produce a thinner disk. As we discussed in the disk radius case, most probably this is an artifact of the limited simulation time combined with faster gas convection for higher density.  
		  The correlation of $H/R$ and $R_{\rm circ} / R_{\rm B}$ is moderate with a correlation coefficient  $C = 0.63$.  
		 The simulation data give 
		 \begin{equation}
		 	0.2 \lesssim H/R \lesssim 0.4,
		 \end{equation}
		 suggesting geometrically slim to thick disks. 
		 
		 In Figure \ref{fig:tau_d_vs_tau_cross_3D} we depict the estimate of disk-formation ignition time $\tau_{\rm d}$ for the 3D case (the time of $\langle v_{\phi} \rangle$ bounce). We get a strong linear correlation for the variables $\tau_{\rm d}$ and $\tau_{\rm cross}$ with coefficient $C = 0.99$. The fit $\tau_{\rm d}/P_{\rm orb} = \lambda_{\tau} \tau_{\rm cross}/P_{\rm orb} + D$ gives 
		 \begin{equation} 
		 	\lambda_{\tau} = 2.9,
		 \end{equation}
		 with $D = 0.004$. The floor value depends on the initial conditions.
		 
		 \section{Conclusion}\label{sec:conclusion}
		 		 
		 	We have suggested a new mechanism of accretion-disk formation around low-mass stellar black holes which move within the cores of gaseous star clusters. A transverse velocity shear is induced by the rotational BH motion, and injects angular momentum into the Bondi sphere as it is advected along the orbit. The injected angular momentum is directed counter to the BH orbit. The Coriolis force opposes this direction and delays the formation of a disk. In the case of eccentric orbits, the variation of angular velocity enhances or opposes counter-rotation in the parts of the orbit where it is accelerated or decelerated, respectively. Still, a disk which rotates counter to the BH orbit eventually forms.
		 	
		 	We have performed 2D and 3D hydrodynamic simulations which verify this picture.
		 	Our simulations suggest that the disk radius is proportional to the circularization radius.
		 	The disk formation time is found to be proportional to the Bondi crossing timescale. For typical star clusters and a BH with mass of $50{\rm M}_{\odot}$, we get a disk-formation timescale of $\sim 0.1 P_{\rm orb}$. This relatively fast ignition of disk formation suggests that the BHs are able to form a disk long before completing an orbit even if they are subject to gravitational perturbations by stellar close encounters or gas turbulence, as investigated in detail by \cite{2026A&A...709A...5R}. Finally, our simulations suggest that the disks are geometrically slim or thick with an aspect ratio $H/R \sim 0.2 - 0.4$ for the range of parameters inspected here.
		 		 
		 	Our code can provide the boundary conditions deep within the Bondi sphere for the investigation of the inner disk with relativistic simulations. Furthermore, future investigations could include the generalization for rotating gas representing, for example, AGN disks, expanding the range of parameters for more diluted systems, and generalizing for the case of a gravitational potential dominated by a central massive object.
		 		 
		 \begin{acknowledgements}
		 	ZR is supported by the European Union's Horizon Europe Research and Innovation Programme under the Marie Sk\l{}odowska-Curie grant agreement No.~101149270--ProtoBH. Numerical calculations have been made possible through a CINECA-INFN agreement, providing access to resources on Galileo100/LEONARDO at CINECA.
		 \end{acknowledgements}
		 		 
		 \bibliography{movingBH_sim}
		 \bibliographystyle{aa_url}	
	 		 	
\end{document}